\documentstyle[12pt,epsf,epsfig]{article}
\def\beq{\begin{equation}}
\def\eeq{\end{equation}}
\def\bea{\begin{eqnarray}}
\def\eea{\end{eqnarray}}
\def\bq{\begin{quote}}
\def\eq{\end{quote}}

\def\gappeq{\mathrel{\rlap {\raise.5ex\hbox{$>$}}
{\lower.5ex\hbox{$\sim$}}}}

\def\lappeq{\mathrel{\rlap{\raise.5ex\hbox{$<$}}
{\lower.5ex\hbox{$\sim$}}}}

\begin{document}
\pagestyle{empty}
\begin{flushright}
{CERN-TH/2000-042}
\end{flushright}
\vspace*{5mm}
\begin{center}
{\bf STRING COSMOLOGY:  THE PRE-BIG BANG SCENARIO} \\
\vspace*{1cm} 
{\bf G. Veneziano}\\
\vspace{0.3cm}
Theoretical Physics Division, CERN \\
CH - 1211 Geneva 23 \\
\vspace*{3cm}  
{\bf ABSTRACT} \\ \end{center}
\vspace*{5mm}
\noindent
A review is attempted of physical motivations, theoretical and
phenomenological aspects, as well as outstanding problems, of the pre-big
bang scenario in string cosmology.
 \vspace*{5cm} 
\noindent 

\noindent
\begin{center}
{\it Lectures given at the Summer School ``The Primordial Universe"}\\
{\it Les Houches, July 1999}
\end{center}

\vspace*{2.5cm}

\begin{flushleft} CERN-TH/2000-042 \\
January 2000
\end{flushleft}
\vfill\eject

\setcounter{page}{1}
\pagestyle{plain}


\def\laq{\raise 0.4ex\hbox{$<$}\kern -0.8em\lower 0.62
ex\hbox{$\sim$}}
\def\gaq{\raise 0.4ex\hbox{$>$}\kern -0.7em\lower 0.62
ex\hbox{$\sim$}}

\def \pa {\partial}
\def \ti {\tilde}
\def \se {{\prime\prime}}
\def \ra {\rightarrow}
\def \la {\lambda}
\def \La {\Lambda}
\def \Da {\Delta}
\def \b {\beta}
\def \a {\alpha}
\def \ap {\alpha^{\prime}}
\def \Ga {\Gamma}
\def \ga {\gamma}
\def \sg {\sigma}
\def \da {\delta}
\def \ep {\epsilon}
\def \r {\rho}
\def \om {\omega}
\def \Om {\Omega}
\def \noi {\noindent}

\def \bp {\dot{\beta}}
\def \bpp {\ddot{\beta}}
\def \fpu {\dot{\phi}}
\def \fpp {\ddot{\phi}}
\def \hp {\dot{h}}
\def \hpp {\ddot{h}}

\def \fb {\overline \phi}
\def \fbp {\dot{\fb}}

\section{Introduction}

These four lectures aim at providing a summary of --and some guidance  
through--
 the existing literature
dealing with the so-called pre-big bang (PBB) scenario, a new  
cosmological model largely based
on the new symmetries underlying  superstring cosmology.
The lectures will be pedagogical in nature and will not presuppose
an advanced knowledge  either of modern inflationary cosmology or of  
superstring/M-theory.
Elements of both will be included in the lectures in order to make them
reasonably self-contained. More exhaustive treatments of pre-big bang
 cosmology are  \cite{Cop} (or will soon be \cite{PRGV}) available  
elsewhere,
while a  homepage on the PBB scenario is being kept
updated on the Web \cite{WEB}.

The four lectures roughly correspond to the four forthcoming sections
and deal, respectively, with:
\begin{itemize}
\item BASIC MOTIVATIONS AND IDEAS
\item  HOW COULD IT  HAVE STARTED?
\item  PHENOMENOLOGICAL CONSEQUENCES
\item  HOW COULD IT HAVE STOPPED?
\end{itemize}
In particular, lecture II (Section 3)  contains a discussion of
 the initial conditions, lecture III (Section 4) discusses the  
phenomenological
virtues and shortcomings of the model,
while lecture IV (Section 5)  deals with the most important open  
theoretical issues.

\section{Basic Motivations and Ideas}

\subsection{Why string cosmology?}

The first question that comes to one's mind when thinking about
cosmology and string theory is: Why bother?
Indeed, even if string/M-theory is the correct
theory of nature,  only its effective (low-energy) quantum field  
theory description
 appears to be relevant to most of the history of our Universe, i.e.
 since a very short time
after the big bang. This is certainly the case for the standard  
(hot-big-bang)
cosmological model, but it is also true for the standard models of inflation,
provided  we confine our
 attention to what happened during the last $70$ e-fold of inflation  
and later
(i.e. to what happened {\it after} our present horizon reached
 the size of the inflationary
Hubble radius). In both instances, one is only confronting situations
in which curvatures are very small with respect to the fundamental
scale of string theory.

On the other hand, both the hot-big bang model and its inflationary variant
 suffer
from  initial condition problems. In the former case,
 these are just the well-known homogeneity and flatness problems that
motivated inflation. In the latter case, although the problems look  
less severe,
it is still a matter of heated discussion
whether or not one should naturally expect a quasi-homogeneous inflaton field
highly displaced from the minimum of its potential to emerge from the  
Planck era.
In either case, the question of how to get physically
appealing initial conditions lies in the realm of Planck-scale quantum  
gravity.

At present, the only candidate for a consistent
 synthesis of general relativity (GR)
and quantum mechanics (QM) is superstring theory (see \cite{Pol} for a  
recent review, as well as \cite{Greene} for a non-specialized  
introduction), or, if we prefer, the mysterious
M-theory that reduces to various superstring theories in appropriate limits.
It  thus seems mandatory to ask whether the above questions on initial
conditions do --or do not-- find an answer within string theory.
Although most string theorists would certainly  agree  with the above  
statements
--this being after all one of the most selling ads for string theory--
many of them would still object to tackling these problems {\it now}.  
The ``excuse"
is that our understanding of string theory, especially at large curvatures, 
is still largely incomplete. Furthermore, most of the recent progress
in non-perturbative string theory has been achieved in the context of  
``vacua" (i.e.
classical solutions to the field equations) that respect a large
number of supersymmetries. By definition, a cosmological background (a  
fortiori
 one that evolves rapidly in time) breaks (albeit spontaneously)  
supersymmetry. This is why
the Planckian regime of cosmology appears to be intractable for the  
time being.

There is however a pleasant surprise. About ten years of
work on string cosmology have led naturally to considering a scenario  --the 
so-called pre-big bang (PBB) scenario-- in which the Universe enjoyed
a long perturbative ``life" {\it before} the big bang.
Starting from an almost trivial
state (asymptotic past triviality, see Section 3),
the Universe would have evolved towards
stronger and stronger curvature and coupling, thereby inflating, until it
entered the non-perturbative phase that replaces the  big bang
singularity of  more standard cosmological models.

The situation is very much reminiscent of QCD and strong interactions.
Perturbative QCD has been very successful in predicting a huge number
of observables for  short-distance-dominated hard processes.
Successes in the non-perturbative, large-distance regime have been meagre, by
comparison: we still lack a definitive proof of confinement, of spontaneous
chiral symmetry breaking, of explicit $U(1)_A$ breaking, etc.
Yet, we do believe that QCD is the correct
description of hadronic physics down to scales of $10^{-15}~{\rm cm}$ or so.
This is largely based on the  belief that large- and short-distance  
physics ``decouple",
e.g. on the assumption that the soft hadronization process does not  
affect certain
infrared-safe quantities computed at the quark--gluon level. Fortunately,
 we did not wait until the confinement problem was solved, to take
QCD seriously!

A very similar attitude will be defended here in the case of string  
cosmology, with
one amusing twist: large- and short-distance physics get somehow  
swapped as we go from
QCD to gravity/cosmology. Figure 1 (from Ref. \cite{GVGvsG})
illustrates this point. The easy regime for gravity is
at large distance/small curvatures;
the tough one turns out to be  the high-curvature regime that replaces
 here the big bang singularity. Yet, we shall argue that
some consequences of string cosmology, those related to scales that were
very large with respect to the string scale in the high-curvature regime,
should not be affected, other than by a trivial kinematical red-shift,  
by the details
of the pre- to post-big bang transition $\dots$ provided, of course,
that such a transition
does indeed take place
(the counterpart to assuming that confinement does occur in QCD).

\begin{figure}
\hglue3.0cm
\epsfig{figure=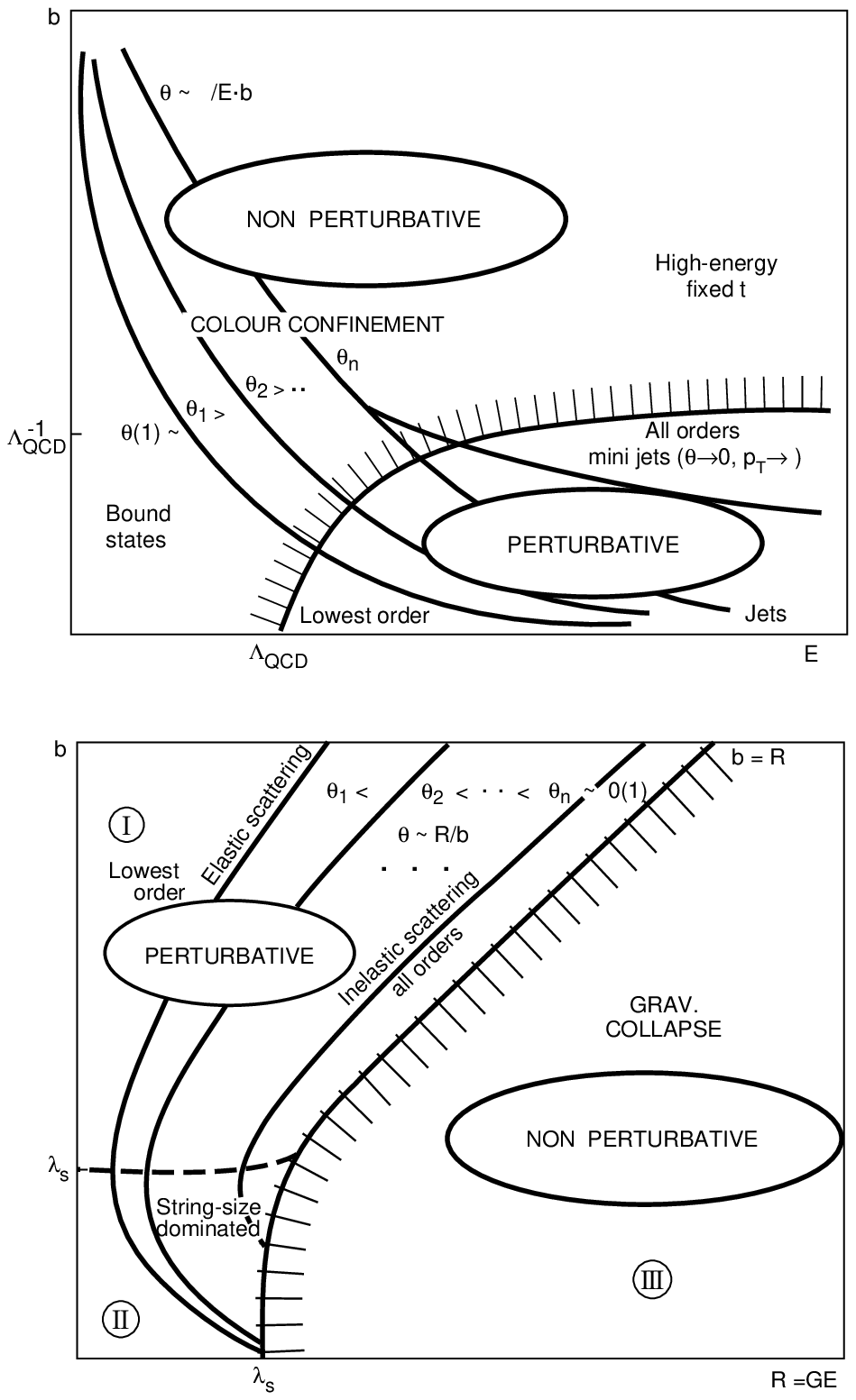,width=10cm}
\begin{center}
Figure 1
\end{center}
\end{figure}

The above reasoning does not imply, of course, that one should not address
 the hard questions {\it now}. On the contrary, the easy part of the  
game will
give precious information on what the relevant hard questions are (for  
cosmology) and
on how to formulate them.
I have already mentioned an example of what I mean: insisting too much  
on (extended) SUSY
vacua appears to be an unacceptable limitation for the problems at hand.
Another example is that of demanding stability of an acceptable string  
vacuum:
we shall see (in Section 4) that inflationary string vacua lead to  
tachyonic,
i.e. to growing rather than to oscillating, modes. Such modes appear  
to horrify
most string theorists; however, they are just what inflationary cosmologists
happily use all the time
in order to generate large-scale structure (LSS), and what PBB cosmology
uses to generate heat and entropy from an initially cold Universe (see  
Section 5).

A completely different criticism of string cosmology
comes  from the cosmology end: for someone accustomed to a data-driven
``bottom-up" approach,
 string cosmology is too much
``top-down". There is certainly a point here. I do not believe that
a good model of cosmology is likely to emerge from theoretical
considerations alone. Input from the data will be
essential in the selection among various
theoretical alternatives. We shall see explicit examples of
what I mean in Section 5. Yet, it appears that
a combination of top-down and bottom-up would be highly desirable.
If past history can teach us something in this respect, the  
construction of the
standard model of particle physics (and of QCD in particular) is a  
perfect example
of a fruitful interplay of theoretically sound ideas and beautiful  
experimental
results. Cosmology today  resembles the particle physics of the  
sixties: interesting new data keep coming in at a high pace, while
 compelling theoretical pillars on which to base our understanding of  
those data are still missing.

As a final remark, let me turn things around and claim that
cosmology could be the only hope that we have for testing
string theory in the foreseeable future by using the cosmos itself as the
largest conceivable accelerator. The cosmological red-shift since the  
big bang
has kindly brought down Planck-scale physics
to a macroscopic scale, thus opening for us a window  on the very early
Universe. As we shall see in Subsection 2.3, even in this respect,
standard and PBB inflation are markedly different.

\subsection{Why/Which inflation?}

The reasons why the standard hot-big-bang model is unsatisfactory
have been repeatedly discussed in the literature. For  details, we refer to
two excellent reviews  \cite{KT}.
Let me briefly summarize here the basic origin of
those difficulties with the simplest  Friedmann--Robertson--Walker  
(FRW) cosmology.

In the FRW framework
the  size of the (now observable)
Universe was about $10^{-2}~ \rm{cm}$ at the start of the classical era,
say at $ t \sim ~{\rm a~ few~ times~} t_P$, where $t_P  \sim 10^{-43}  
\; \rm{s}$
is the so-called Planck time.
 This is of course a very tiny Universe  w.r.t. its present size
($\sim 10^{28}~ \rm{cm}$), yet it is huge w.r.t. the horizon (the distance
travelled by light) at that time, i.e.
to $l_P = c t_P \sim 10^{-33}~ \rm{cm}$.
In other words, a few Planck times after the big bang,
 our observable Universe was much too large! It consisted
of $(10^{30})^3 = 10^{90}$ Planckian-size, causally disconnected regions.
There had not been, since the beginning, enough time
 for the Universe to become homogeneous (e.g. to thermalize)
over its entire size.
Also, soon after $t=t_P$, the Universe was characterized by a huge hierarchy
between its Hubble radius on one side and its spatial-curvature
radius  on the other. The relative factor of (at least) $10^{30}$ appears
 as an incredible amount of fine-tuning on the initial state of the Universe,
corresponding to a huge asymmetry between time and space derivatives.  
Was this asymmetry
really there? And, if so, can it be explained in any, more natural way?

It should be stressed that, while the above unexplained ratio becomes
larger and larger as we approach the Planck time (and would go to infinity
at $t=0$ if we could trust the equations throughout), it represents  
the  ratio of
two {\it classical} length scales. It so happens that one of the two lengths
becomes the (quantum) Planck scale {\it at $t=t_P$},  but the ratio is  
still  huge
at much later times when both scales have nothing to do with (and are  
much larger than)
 $t_P$. This comment will be very relevant to the discussion of fine-tuning
issues given in Subsection 3.6.

It is well known that a generic way to wash out inhomogeneities and spatial
curvature consists in introducing, in the history of the Universe, a long
period of accelerated expansion, called inflation \cite{KT}.
This still leaves two alternatives: either the Universe was generic
at the big bang and became flat and smooth
because of a long {\it post}-bangian inflationary phase;
 or it was
already flat and smooth at the big bang as a result of
 a long {\it pre}-bangian inflationary phase.

Assuming, dogmatically, that the Universe (and time itself) started at  
the big bang,
 leaves
 only the first alternative. However, that solution has its own
problems, in particular those of
fine-tuned initial conditions and inflaton potentials. Besides, it is  
quite difficult \cite{BS}
to base standard inflation in the only known candidate theory of  
quantum gravity,
superstring theory. Rather, as we shall argue in a moment, superstring
 theory gives strong hints in favour of the
second (pre-big bang) possibility through two of its very basic properties, 
the first in relation
to its short-distance behaviour, the second from its modifications of  
GR even at large distance.

\subsection{Superstring-inspired cosmology}

As just mentioned, two  classes of properties of string theory
are relevant for cosmology. Let us discuss them in turn.

A) Short-distance properties
\vskip 5mm

Since the classical (Nambu--Goto)  action of a string
is proportional to the area $A$ of the surface it sweeps, its  
quantization must introduce
a quantum of length $\lambda_s$ through:
\begin{equation}
S/\hbar =  A/\lambda_s^2\; .
\end{equation}
This fundamental length, replacing Planck's constant in quantum string  
theory \cite{GVFC},
plays the role of a minimal observable length, of an ultraviolet cut-off.
Thus, in string theory,  physical quantities are expected to be bound  
by appropriate
powers of $\lambda_s$, e.g.
\begin{eqnarray}
H^2 \sim R \sim G\rho < \lambda_s^{-2} \nonumber \\
k_B T/\hbar < c \lambda_s^{-1} \nonumber \\
R_{comp} > \lambda_s \; .
\end{eqnarray}
 In other words, in quantum
string theory,  relativistic quantum mechanics should solve the singularity
problems in much  the same way as
non-relativistic quantum mechanics solved the singularity
problem of the hydrogen atom by keeping the electron and the proton a  
finite distance
apart. By the same token, string theory
 gives us a rationale for asking  daring questions such as:
What was there before the big bang? Certainly, in no other present  
theory can
such a question
be meaningfully asked.

\vskip 5mm
B) Large-distance properties
\vskip 5mm

Even at large distance (low-energy, small curvatures), superstring  
theory does {\it not}
 automatically give Einstein's GR. Rather, it leads to a
scalar--tensor theory of the JBD variety. The new scalar  
particle/field $\phi$,
the so-called dilaton, is unavoidable in string theory, and gets
reinterpreted as the radius of a new dimension of space in so-called  
M-theory \cite{M}.
By supersymmetry, the dilaton is massless to all orders in  
perturbation theory,
i.e. as long as supersymmetry remains unbroken. This raises the question:
Is the dilaton a problem or an opportunity? My answer is that it could  
be both;
and while we can try to avoid its potential dangers, we may try to use  
some of
its properties to our advantage \dots~ Let me discuss how.

In string theory, $\phi$ controls the strength of all forces  
\cite{Witten}, gravitational and gauge alike.
One finds, typically:
\begin{equation}
 l_P^2 /\lambda_s^2  \sim \alpha_{gauge} \sim e^{\phi} \; ,
\label{VEV}
\end{equation}
showing the basic unification of all forces in string theory and the  
fact that, in
our conventions,
 the weak-coupling region coincides with $\phi \ll -1$.
In order not to contradict precision tests of the
equivalence principle, and of the constancy of the gauge and gravitational
couplings in the ``recent" past,
we require \cite{TV} the dilaton to have a mass (see, however,
 \cite{DP} for an amusing alternative)
 and to be frozen at the bottom of
 its own potential
{\it today}. This does not exclude, however, the possibility of the  
dilaton having
evolved cosmologically (after all, the metric did!) within the weak coupling
region where it was practically massless. The amazing (yet simple)  
observation \cite{GV1}
is that, by so doing, the dilaton may have inflated the Universe!

A simplified argument, which, although not completely accurate,  
captures the essential
physical point, consists in writing the Friedmann equation (for a  
spatially flat Universe):
\begin{equation}
3 H^2 = 8 \pi G \rho\; ,
\label{FRE}
\end{equation}
 and in noticing that a growing dilaton (meaning through (\ref{VEV}) a  
growing $G$)
can drive the growth of $H$ even if the energy density of standard  
matter decreases
 in an expanding Universe. This new kind of inflation (characterized by 
growing $H$ and $\phi$) has been
termed dilaton-driven inflation (DDI). The basic idea of pre-big bang
cosmology \cite{GV1,MG1,MG2,MG3}
is thus illustrated in Fig. 2: the dilaton  started at very
large negative values (where it was practically massless), ran over a  
potential hill, and finally reached,
sometime in our recent past, its final destination at the bottom of
its potential ($\phi = \phi_0$). Incidentally, as shown in Fig. 2,
the dilaton of string theory can easily
roll-up ---rather than down--- potential hills, as a consequence of  
its non-standard
coupling to gravity.

\begin{figure}
\hglue3.0cm
\epsfig{figure=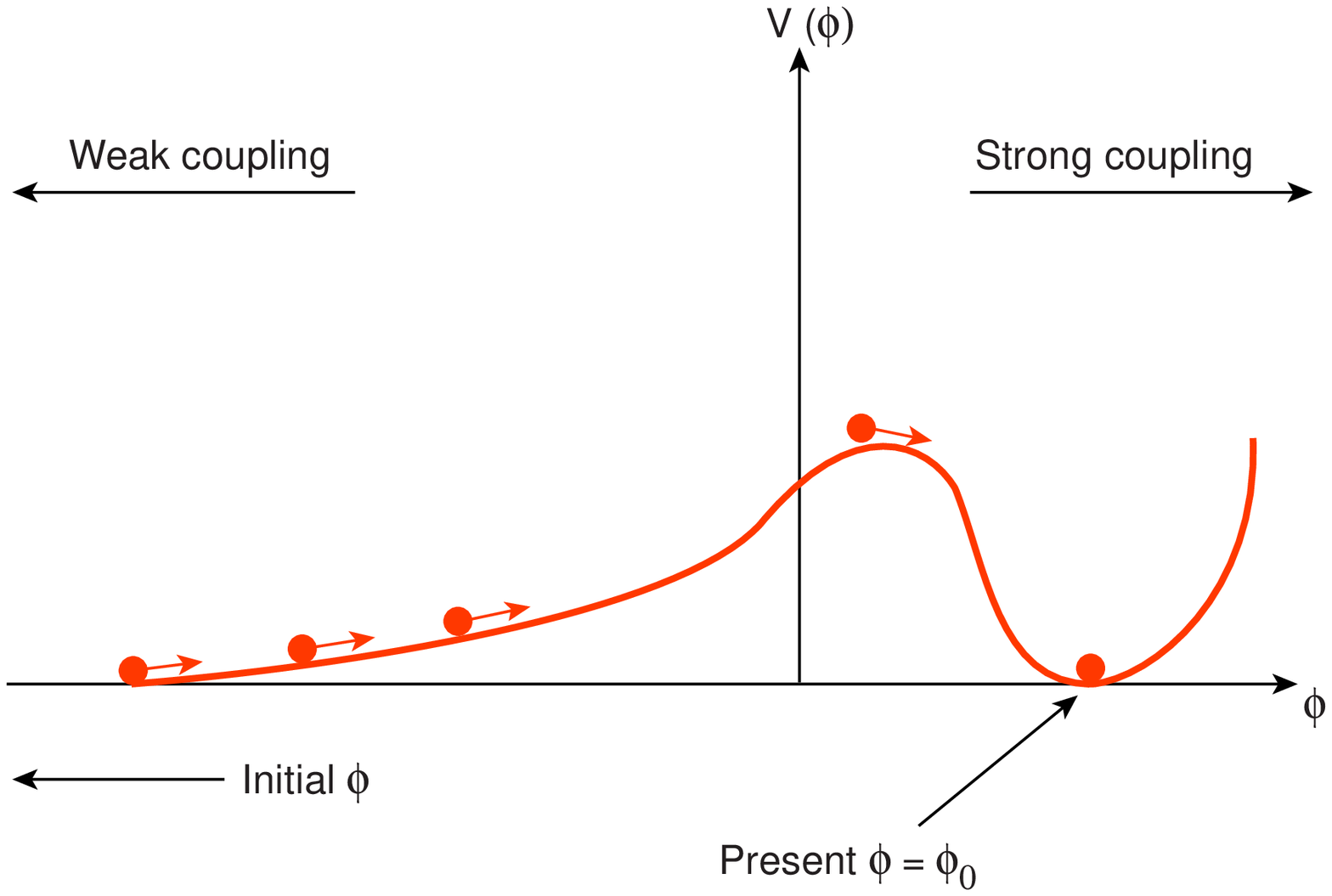,width=10cm}
\begin{center}
Figure 2
\end{center}
\end{figure}

DDI is not just possible. It exists as a class of (lowest-order)  
cosmological solutions
thanks to the duality symmetries of string cosmology \cite{GV1}, \cite{SFD},
 \cite{ODD}. Under
a prototype example of these symmetries, the
so-called scale-factor duality (SFD) \cite{GV1}, \cite{SFD}, a FRW cosmology
evolving (at lowest order in derivatives) from a singularity in the past 
is mapped into a DDI cosmology going towards a singularity in the  
future. Of course,
the lowest order approximation breaks down before either singularity  
is reached.
A (stringy) moment away from their respective
singularities, these two branches can easily be joined smoothly to give 
a single non-singular cosmology, at least mathematically.
Leaving aside this issue for the moment (see Section 5 for more discussion),
let us go back to DDI.
 Since such a phase is characterized by growing coupling and curvature, it
 must itself have originated from
 a regime in which both quantities were very small. We take this as
 the main lesson/hint  to be
learned from low-energy string theory by raising  it to the level
of a new cosmological principle, that  of ``Asymptotic Past Triviality", 
to be discussed in the next Lecture.

\subsection{Explicit solutions}

Many explicit  exact PBB-type solutions to the low-energy effective action
 equations have been constructed and discussed in the literature. For  
an excellent review,
see \cite{Cop}. Exact solutions can only be obtained in the presence
of symmetries (isometries) and, although they are heuristically very  
important,
they are too special from the point of view of an inflationary  
cosmology, which,
as such, should not
accept fine-tuned
initial conditions. This is why we shall not go into an exhaustive discussion
of explicit solutions here. Instead, in Section 3, we will adress  
the general problem
of the evolution of asymptotically trivial initial data.

Here we shall limit our attention to the simplest Bianchi I-type solutions
and to their quasi-homogeneous generalizations, after recalling that many
 more solutions can be obtained from the former by using the  
non-compact $O(d,d)$
symmetry of the low-energy string-cosmology equations \cite{ODD}
when the Kalb--Ramond (KR) field $B_{\mu\nu}$ is turned on, or by  
S-duality transformations (see e.g.
\cite{Cop})
generating a homogeneous axion field (related to $B_{\mu\nu}$ by yet  
another duality
transformation).

The generic homogeneous Bianchi I solution with $B_{\mu\nu}=0$
reads, for $t<0$,
\begin{eqnarray}
d s ^2 &=&  - dt^2 + \sum_i (-t)^{2 \alpha_i} dx^i dx^i  \;,  \nonumber \\
~~ \phi &=& - (1 - \sum_i \alpha_i) {\rm log} (-t) \nonumber \\
1 &=& \sum_i \alpha_i^2 \; .
\label{Kasner}
\end{eqnarray}
i.e. represents a generalization of the well-known Kasner solutions
 (see e.g. \cite{Zeldovich}) in which
one of the two Kasner constraints (the one linear in the $\alpha_i$)  
is replaced
by the equation giving the time dependence of $\phi$ ($\phi$ is  
absent, or constant,
for Kasner, hence the second constraint).

Note that, unlike Kasner's, (\ref{Kasner}) allows for isotropic solutions
($\alpha_i = \pm 1/ \sqrt{d}$ for all $i$). Also, the quadratic Kasner  
constraint
automatically has $2^d$ SFD-related branches, obtained by changing the sign
of any subset of the $\alpha's$.
Also note that the so-called shifted dilaton defined by:
\begin{equation}
\bar{\phi} = \phi - \frac {1} {2} {\rm log}~({\rm det}~ g_{ij})  \; ,
\label{sdildef}
\end{equation}
which is invariant under the full $O(d,d)$ group, is always given by:
\begin{equation}
\bar{\phi} = - {\rm log} (-t)  \; .
\label{sdileqn}
\end{equation}

A quasi-homogeneous generalization of (\ref{Kasner}) was first  
discussed in \cite{inhom}
(see also \cite{BMUV1})
and reads:
\begin{eqnarray}
d s ^2 &=&  - dt^2 + \sum_a e_i^a(x)~ e_j^a(x) (-t)^{2 \alpha_a(x)}  
dx^i dx^j  \;,  \nonumber \\
~~ \phi &=& - (1 - \sum_i \alpha_i(x)) {\rm log} (-t) \nonumber \\
1 &=& \sum_i \alpha_i^2(x) \; ~~ ~~, ~~~~ t < 0 \; ,
\label{inhKasner}
\end{eqnarray}
where $x$ stands for the space coordinates.
Equation (\ref{inhKasner}) can be shown to be a generic asymptotic
 solution of the full PDEs near the $t=0$
singularity where spatial gradients become
less and less important w.r.t. time derivatives, justifying
the validity of the so-called gradient expansion \cite{GE}.
 Note that Eq. (\ref{sdileqn}) is not modified in the quasi-homogeneous
solutions.  Besides allowing isotropic cosmologies in the homogeneous
case, the presence of the dilaton also removes the necessity of a  
chaotic (BKL-type \cite{BKL})
behaviour  near the singularity \cite{BD}.

\subsection{Phase diagrams and Penrose-style overview}

It is useful to visualize the PBB scenario with the help of some diagrams.
Since the actual phase space of the model is multidimensional, each of these
diagrams necessarily represents just a cross section of the complete picture.

A very commonly used diagram (Fig. 3) is the flow-diagram in the  
$\dot{\bar{\phi}},H$ plane
(time being just a parameter along the flow lines). Since, at lowest order,
$\ddot{\bar{\phi}} \ge 0$, the flow is always from left to right near  
the origin.
The four straight lines
represent the four (isotropic for simplicity) solutions connected by  
SFD and time-reversal.
The product of the two transformations represents the physically  
interesting case,
since it maps ordinary decelerating FRW cosmology (top left) to  
dilaton-driven
inflation (top right). Clearly, our scenario needs a high-curvature phase
during which the left-to-right flow is inverted (as shown by the dotted line
joining the two perturbative branches). This can only happen as the result
of higher-order corrections (see Section 5).

\begin{figure}
\hglue3.0cm
\epsfig{figure=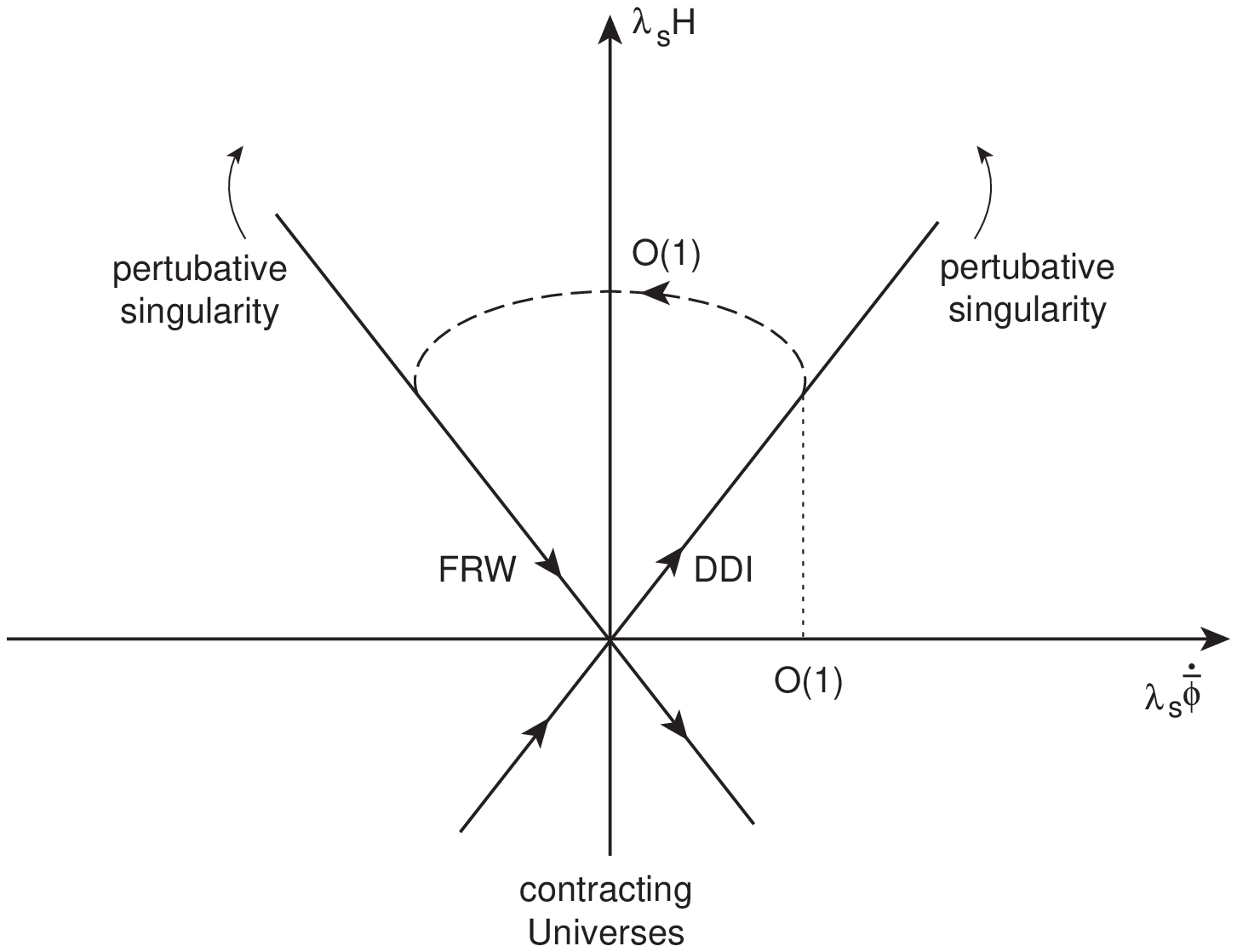,width=10cm}
\begin{center}
Figure 3
\end{center}
\end{figure}

A second useful diagram (Fig. 4) is the $e^{\phi}, H$ plot, i.e.
the curvature (energy) coupling
plane. The fully perturbative domain (where evolution starts
according to the APT postulate) lies, in a log-log plot, to the far  
left-bottom corner.
Sticking again, for simplicity,
 to the isotropic case, DDI evolution is represented by parallel lines 
distinguished by different initial values of the dilaton (i.e. of the  
coupling).
It is clear that all these solutions run, eventually,
 into strong curvature or strong coupling (shown as thick solid lines),
which one is hit first being determined by the above-mentioned initial  
coupling. A discussion of what might happen afterwards is given in  
Section 5.

\begin{figure}
\hglue3.0cm
\epsfig{figure=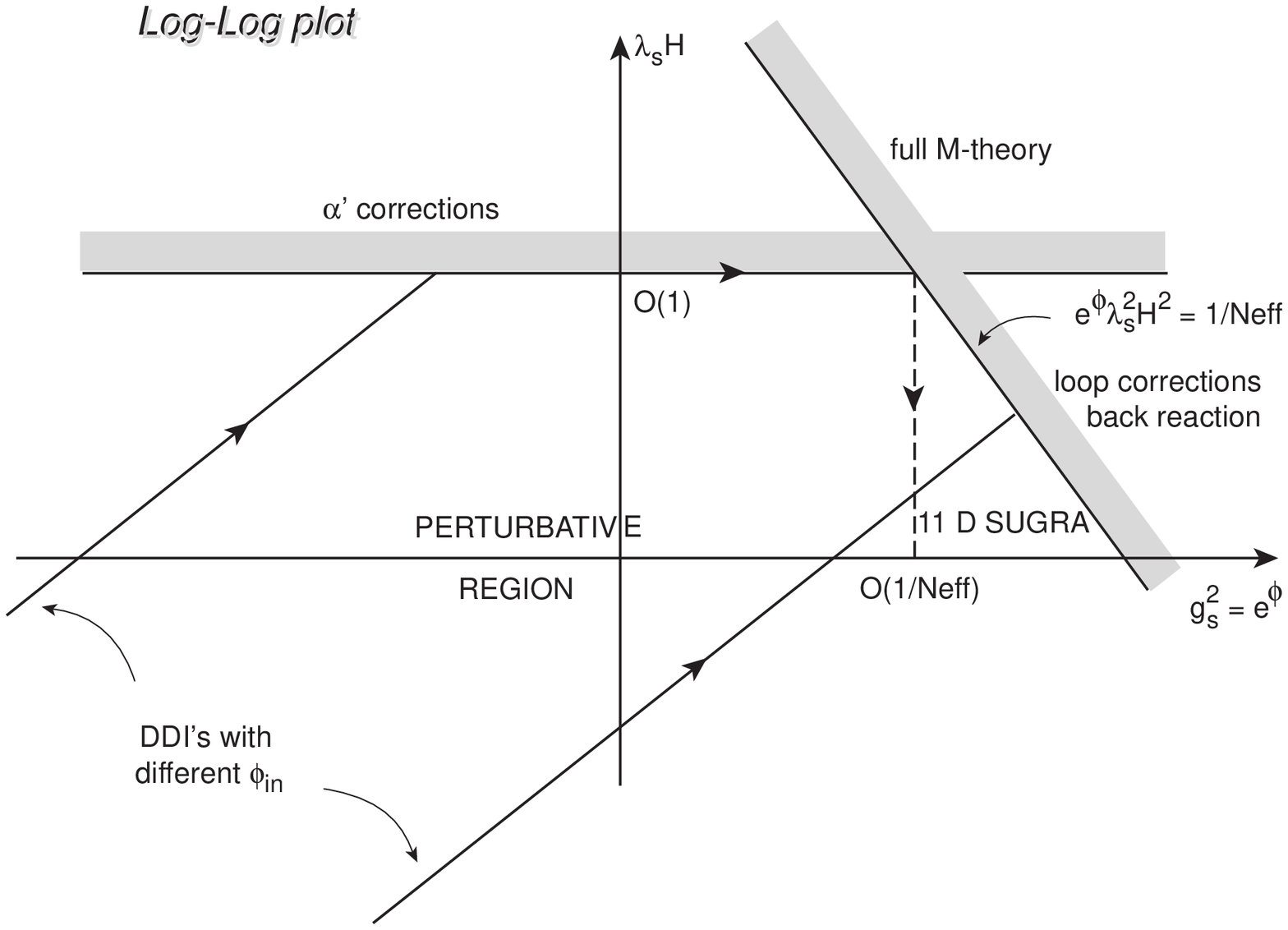,width=10cm}
\begin{center}
Figure 4
\end{center}
\end{figure}

As a third possibility, let us
 use a Carter--Penrose style plot \cite{CP} (Fig. 5) to represent,
on a finite piece of
paper, the entire evolution of the Universe. Unlike in ordinary
cosmology, where the CP diagram is truncated by the (space-like) hypersurface
of the big-bang singularity, here the whole CP diagram, going from  
past to future
time-like and null infinities, is physically meaningful because of our  
assumption
that finite-string-size effects remove the
big bang singularity. This diagram
will be discussed and used in the following sections.

\begin{figure}
\hglue3.0cm
\epsfig{figure=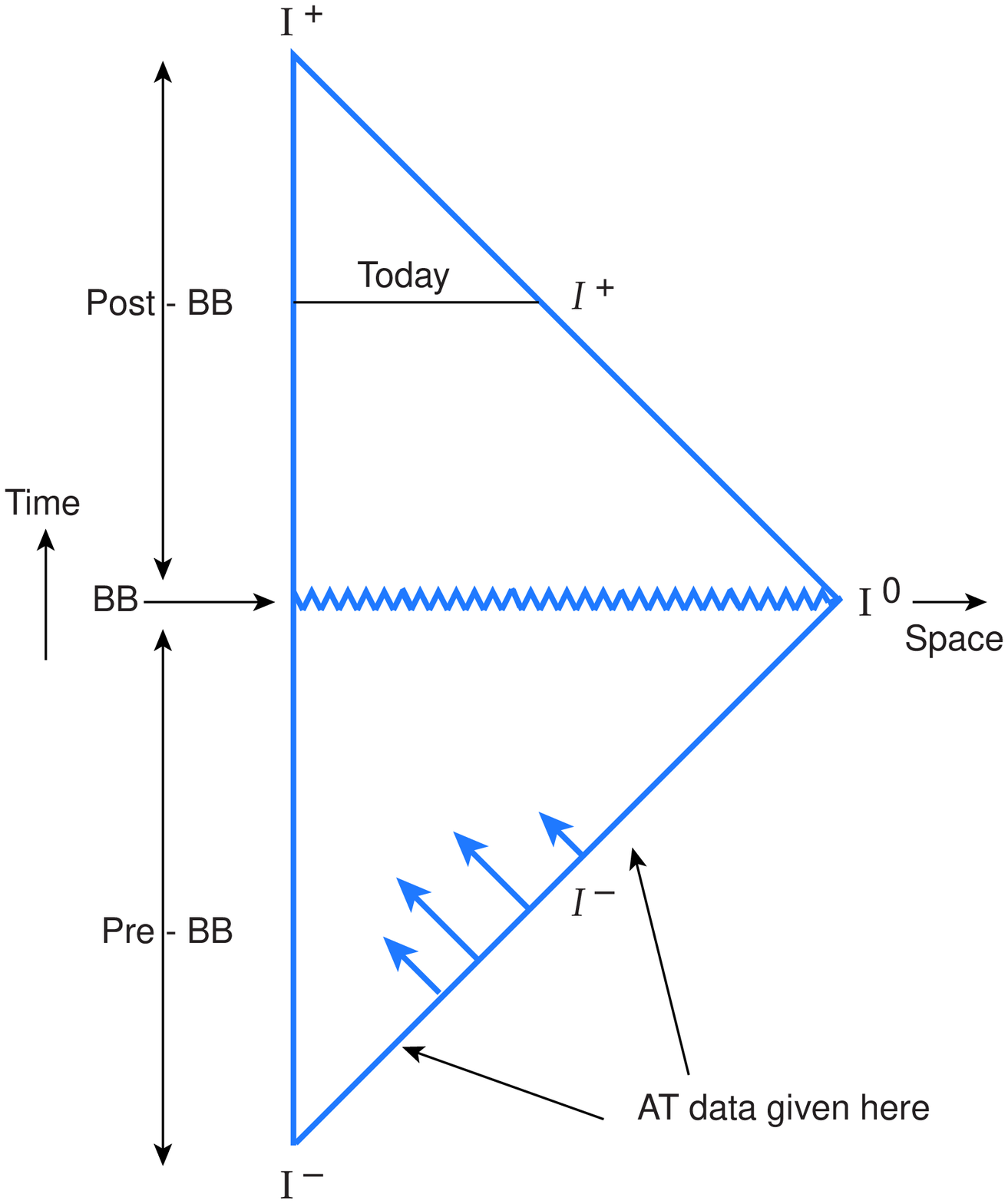,width=10cm}
\begin{center}
Figure 5
\end{center}
\end{figure}

\begin{figure}
\hglue3.0cm
\epsfig{figure=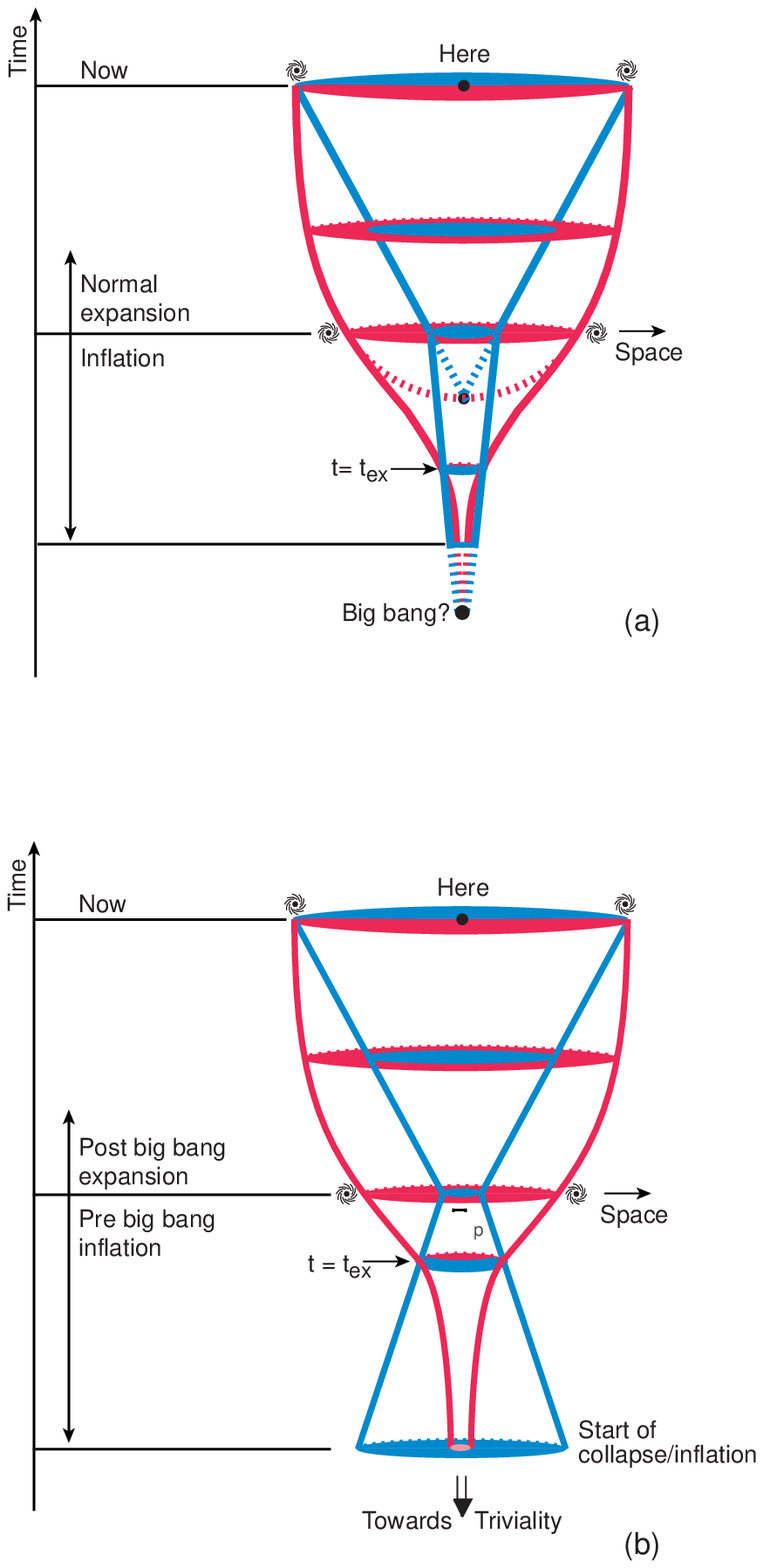,width=10cm}
\begin{center}
Figure 6\\
(a) Standard inflation's wineglass; (b) Pre-Big Bang's wineglass
\end{center}
\end{figure}

Finally, let us represent the basic difference between
the standard inflation scenario and that of
PBB cosmology by plotting, for each cosmological model, the
Hubble horizon ($H^{-1}$) and the physical
scale that coincides with it today, as functions of cosmic time.
This gives rise to two ``wine glasses" (Fig. 6), which are very similar
in their upper parts (corresponding to recent epochs) but differ
markedly at very early times. The most salient difference appears
in the early behaviour of the Hubble horizon, an increasing function of time
in the standard inflation, a decreasing one in the PBB case.
The figure allows me to stress one phenomenological advantage
of PBB inflation:
Planck- (or string)-scale physics, being no longer washed out by a  
long, subsequent
inflationary phase, becomes  accessible to present (or near-future)  
experiments
at the millimetre ($100$ GHz) scale. At the same time, larger-scale  
experiments
(such as those
on small-angle CMB  anisotropies) will test (sub-Planckian-energy)  
physics during the
pre-bangian phase. By contrast, as we have already mentioned in the  
Introduction,
in standard
inflation large-scale data probe the Universe as it was
 seventy e-folds or so before the end of
inflation, while shorter scales tells us about more recent epochs.
 Since we know that,
seventy e-folds before the end of inflation,
  $H_{infl}$ was less than $10^{-5} M_P$ (or else  excessive
 large scale anisotropies are created, see Section 4), and
that such a scale slowly {\it decreases} during
(slow-roll) inflation, it is clear that, according to standard
inflation, physics at energies larger than $10^{-5} M_P$
remains unaccessible.

\section{How could it  have started?}

\subsection{Generic asymptotically-trivial past}

We have already mentioned that, in standard non-inflationary  
cosmology, initial
conditions have to be fine-tuned to incredible accuracy in the
far past (i.e. at $t \sim t_P \sim 10^{-43}$ s).
What does this fine-tuning problem look like if we accept hints from
scale-factor duality and assume asymptotically trivial, yet generic, initial
conditions?

The concept of asymptotic past triviality (APT)
 is quite similar to that of ``asymptotic flatness",
familiar from general relativity \cite{AF}. The main differences  
consist in making
 only  assumptions
concerning the asymptotic past (rather than future or space-like infinity)
 and in the additional presence
of the dilaton. It seems physically (and philosophically)
satisfactory to identify the beginning with simplicity (see e.g. the  
entropy-related
 arguments
given in Subsection 5.7). What could be  simpler than a trivial,
empty and flat Universe? Nothing, of course! The problem is that such  
a Universe,
besides being uninteresting, is also
non-generic. By contrast, asymptotically flat/trivial
 Universes are initially simple, yet generic,
 in a precise mathematical sense that we shall now discuss.

From the point of view of space-time (taken here, for simplicity, to  
be $(3+1)$-dimensional)
the generic solution depends upon four arbitray functions of three  
coordinates \cite{LL}
related to the metric, plus two more each for the dilaton and the KR  
field $B_{\mu\nu}$.
Amusingly, there is an exact correspondence between this  
``target-space" counting and
a ``world-sheet" counting. In the latter, those eight arbitrary functions 
correspond to eight arbitrary functions of three-momentum entering the most
general physical (i.e. on shell) vertex operator describing gravitons,  
dilatons,
and the KR field (which,
in four dimensions, is equivalent to a pseudoscalar, the KR axion).
 We will see in Subsection 3.4 how these arbitrary functions appear  
in the asymptotic
expansion of our fields.

Can a very rich and complicated Universe, like our own,
 emerge from such extremely simple initial conditions?
  This would look much like a miracle. However, as I shall
  argue below, this is precisely what
should be expected, owing to well-known classical and  quantum gravitational 
instabilities.

\subsection{The asymptotic past's effective action and different  
(conformal) frames}

The APT postulate implies that the early-time evolution of the Universe
can be described in terms of the low-energy tree-level action
of string theory. Taking a generic closed superstring theory, this reads:
\beq
\Gamma_{eff} =  \lambda_s^{1-d}\int\,d^{d+1}x\,\sqrt{|g|}\,e^{-\phi}
\,\left(R + g^{\mu\nu}\partial_{\mu}\phi\partial_{\nu}\phi - \frac{1}{12}
 (dB)^2 - 2~ \Lambda \right) ~~,
\label{21}
\eeq
where $dB$ is the (three-form) field strength associated with $B_{\mu\nu}$.

A further simplification comes from assuming to be dealing with so-called
critical superstring theory, the case in which the tree-level (and actually
the all-order perturbative) cosmological constant $\Lambda$ vanishes.  
This requires
a total of $D= 10$ space-time dimensions. If $D \ne 10$ there will be  
an effective
cosmological constant $O(\lambda_s^{-2})$ preventing any low-curvature
solution of the field equations to exist. A similar conclusion is reached
if we consider critical, but non-supersymmetric, string theories (see  
Subsection 3.3).

Equation (\ref{21}) receives corrections when curvatures become  
$O(\lambda_s^{-2})$
{\it or} when the coupling $e^{\phi}$ becomes $O(1)$. If  such  
corrections are both negligible,
it sometimes becomes  useful to perform a change of variable by going to the
so-called Einstein frame (not to be confused with different frames
 in GR).
This is done by defining:
\begin{equation}
g_{\mu\nu} = g^{(E)}_{\mu\nu} e^{\frac{2}{d-1}(\phi - \phi_0)} \; .
\label{ESF}
\end{equation}

It is relatively easy to rewrite the action (\ref{21}) using the  
Einstein metric.
The result is simply:
\beq
\Gamma_{eff}^{E} =  l_P^{1-d} \int\,d^{d+1}x\,\sqrt{|g^{(E)}|}\;
\,\left(R - \frac{1}{d-1} \partial_{\mu}\phi\partial^{\mu}\phi - \frac{1}{12}
 e^{- \frac{4}{d-1} \phi} (dB)^2  \right) \; ,
\label{24}
\eeq
where $ l_P^{d-1} = e^{\phi_0} \lambda_s^{d-1}$ is the present value of the
Planck length.

Although the use of the Einstein frame could simplify some calculations, and
we shall see examples of this below, it should be kept in mind that
the form of the corrections is no longer so simple. For instance,  
higher-derivative
corrections become important when the Einstein-frame curvature is
$O( l_P^{-2} e^{-\frac{2}{d-1} \phi} = \lambda_s^{-2} )$, i.e. reaches a 
dilaton-dependent critical value.
Similarly, having a constant Newton ``constant" in this frame is a  
mere illusion
because (even tree-level) string masses do now depend upon $\phi$.
For these reasons, although physical results are frame-independent, we  
shall always
describe them  with reference to the original string-frame metric
in which the stringh length $\lambda_s$ is constant.

Let us finally remark that the two frames have  been made to coincide
today, with the dilaton fixed at its present value $\phi_0$. Similarly, 
the assumption of APT would also allow the identification of the two frames
in the far past, since the dilaton approaches a constant as $t  
\rightarrow - \infty$.
 However, the two Einstein frames that coincide with the string frame at 
$t = \pm \infty$ differ  from each other by an enormous conformal  
factor, i.e.
by a huge blowing-up
 of all physical scales.

\subsection{ Classical asymptotic symmetries: the importance of SUSY}

The classical equations that follow from varying (\ref{21}) or (\ref{24}), 
besides being generally covariant,
are also invariant under a two-parameter group of (global)  
transformations acting as follows:

\begin{eqnarray}
\phi &\rightarrow&  \phi + c  \;,  \nonumber \\
g_{\mu\nu} &\rightarrow&  \lambda^2 g_{\mu\nu}\; .
\label{symmetries}
\end{eqnarray}
Indeed  (\ref{21}), (\ref{24}) are simply rescaled by a constant factor
under this group. These
two symmetries depend crucially on the validity of the tree-level  
low-energy approximation
and on the absence of a cosmological constant. Loop corrections
clearly spoil invariance under dilaton shifts, while lower derivatives  
(a cosmological
constant) or higher derivatives ($\alpha'$) corrections spoil  
invariance under
a rescaling of the metric. Note that, using general covariance, the latter
symmetry is equivalent to an overall rescaling of all the coordinates.
The relevance of the two classical symmetries on the issue of
fine-tuning will become obvious in the next two subsections.

The importance of dealing with critical
superstring theory now  becomes evident:
 if one would consider non-supersymmetric string theories, a  
cosmological constant
would almost certainly be generated at some finite order of the loop
expansion:  this would change completely the large-distance properties and
spoil the symmetries of the field equations.

\subsection{Dilaton-driven inflation as gravitational collapse}

 For simplicity, we will only illustrate here the simplest case of gravi-dilaton
 system already compactified to four space-time dimensions.  Through the field redefinition (\ref{ESF}),
our problem is  reduced to the study of a massless scalar field
minimally coupled to gravity.
It is well known that such a form of matter cannot give inflation  
(since it has
positive pressure). Instead, it can easily lead to gravitational  
collapse (GC).
Thus, in the Einstein frame, the problem becomes that of
finding out  under which conditions  gravitational collapse  occurs if 
asymptotically-trivial initial data are assigned.
Gravitational collapse  usually
means that the (Einstein) metric (hence the volume of 3-space)
 shrinks to zero at
a space-like singularity. However, typically,
the dilaton blows up at that same singularity. Given the relation (\ref{ESF})
between the Einstein and the (physical) string metric, we can easily  
imagine that
the latter blows up near the singularity, as implied by DDI.

How generically does GC happen? Let us recall
the singularity theorems of Hawking and Penrose \cite{HP}, which state  
that, under some
general assumptions, singularities are inescapable in GR.
Looking at the validity of those assumptions in the case at hand, one  
finds that
 all but one are automatically satisfied. The only condition to be
imposed is the existence of a closed trapped surface (CTS) (a closed surface
 from which  future light cones lie entirely in the region inside the  
surface).
Rigorous results \cite{Chr} show that this condition cannot be waived: 
sufficiently weak initial data
do not lead to closed trapped surfaces, to collapse, or to  
singularities. Sufficiently
strong initial data do. But where is the border-line? This is not  
known in general,
but precise criteria do exist for particularly symmetric space-times,  
e.g. for those
endowed with spherical symmetry (see Subsection 3.6).

However, no matter what the general collapse/singularity criterion will 
eventually turn out to
be, we do know, from the classical symmetries  described in the  
previous subsection,
 that such a criterion cannot depend
\begin{itemize}
\item  on an over-all additive constant in $\phi$, or
\item   on an over-all multiplicative factor in $g_{\mu\nu}$.
\end{itemize}

A characterization of APT initial data can be made \cite{BDV}  
following the pioneering work
 \cite{AF} of
Bondi, Sachs, Penrose, and others. Since our initial quanta are  
assumed to consist
of massless
gravitons and dilatons, their past infinity is null: it is the famous  
${\cal I}^-$ of
the Penrose diagram (Fig. 5). APT means that dilaton and metric can be  
expanded
near ${\cal I}^-$ in inverse powers of $r \rightarrow \infty$, while  
advanced time $v$
and two angular variables, $\theta$ and $\varphi$, are kept fixed. We  
shall thus write:
\beq
\phi (x^{\lambda}) = \phi_0 + \frac{f(v,\theta , \varphi)}{r} + {o} \left( 
\frac{1}{r} \right) \, ,
\label{eqn2.4}
\eeq
\beq
g_{\mu \nu} (x^{\lambda}) = \eta_{\mu \nu} + \frac{f_{\mu \nu} (v,\theta , 
\varphi)}{r} + {o} \left( \frac{1}{r} \right) \, .
\label{eqn2.5}
\eeq
The null wave data on ${\cal I}^-$ are: the asymptotic dilatonic wave  
form $f
(v,\theta , \varphi)$, and  two polarization components, $f_+ (v,\theta , 
\varphi)$ and $f_{\times} (v,\theta , \varphi)$, of the asymptotic  
gravitational
wave form $f_{\mu \nu} (v,\theta , \varphi)$, whose other components can be
gauged away.
The three functions $f$,
$f_+$, $f_{\times}$ of $v,\theta , \varphi$ are equivalent to six functions 
 of $r,\theta , \varphi$ with $r \geq 0$, because
the advanced time $v$ ranges over the full line $(-\infty , +\infty)$.  
This is
how the six arbitrary functions of the generic solution to the gravi-dilaton
system are recovered.

Of particular interest here are the so-called News functions, simply
given by
\beq
N (v,\theta , \varphi) \equiv \partial_v \, f (v,\theta , \varphi) \ , \ N_+ 
\equiv \partial_v \, f_+ \ , \ N_{\times} \equiv \partial_v \,  
f_{\times} \, ,
\label{eqn2.8}
\eeq
and the ``Bondi mass'' given by:
\bea
M_-(v) &=& \frac{1}{4\pi} \int d^2\Omega M_-(v, \theta, \varphi)~~,    
\nonumber \\
g_{vv} &=& - \left( 1- {2 M_-(v,\theta , \varphi) \over r}\right) +  
{o} \left( \frac{1}{r} \right)\; .
\label{Bondi}
\eea
The Bondi mass and the News are connected by
the energy--momentum conservation equation, which tells us that the  
advanced-time
derivative of $M_-(v)$
is positive-semidefinite and related to incoming energy fluxes  
controlled by the News:
\beq
dM_- (v)/dv  = \frac{1}{4} \, \int d^2\Omega \left(
N^2 +  \, N_+^2 +  \,
N_{\times}^2 \right)\, .
\label{eqn2.7}
\eeq

The physical meaning of $M_-(v)$ is that it represents the energy brought
into the system (by massless sources) by advanced time $v$.
In the same spirit one can define the Bondi mass $M_+(u)$ at future null
infinity ${\cal I}^+$.
It represents the energy still present in the system at  retarded time $u$.
If only massless sources are present, the so-called ADM mass is given by 
\begin {equation}
M_-(+\infty) = M_+(-\infty) = M_{ADM} \; ,
\end{equation}
while $M_-(-\infty) = 0$, and  $M_+(+\infty) = M_C$ represents the mass that
 has not been radiated away
even after waiting an infinite time, i.e. the mass that
 underwent gravitational collapse \cite{Chr87}. Collapse (resp. no-collapse)
 criteria thus aim at establishing
under which initial conditions one expects to find  $M_C > 0$ (resp.  
$M_C =0$).

Since, as we shall see in the particular case of spherical symmetry,
collapse criteria i) do not involve any particularly large number,
and ii) do not contain any intrinsic scale but just dimesionless  
ratios of various
classical scales, we expect i) gravitational collapse to be quite a  
generic phenomenon
and ii) that nothing, at the level
of our approximations, will be able to fix either the size of the horizon
 or the value of $\phi$ at the onset of collapse.
Generically, and quite randomly and chaotically,  some regions of space
 will undergo gravitational
collapse,  will form horizons and singularities therein. When this is  
translated
into the string frame,
 the region of space-time within the horizon undergoes a period of DDI
in which both the initial value of the Hubble parameter and that of  
$\phi$ are left
arbitrary. In the next subsection we shall see that such arbitrariness 
provides an answer to the fine-tuning allegations that
have been recently moved \cite{TW} to the PBB scenario.
This section will be concluded with a discussion
of how more precisely  the case of spherical symmetry can be dealt with.

\subsection{Fine-tuning issues}

The two arbitrary  parameters discussed in the previous subsection
are very important, since they
determine the range of validity of our description. In fact, since  
both curvature and
coupling increase during  DDI,  the low-energy and/or tree-level
description is bound to break down at some point. The smaller the  
initial Hubble parameter (i.e. the larger
the initial horizon size) and the smaller the initial coupling, the  
longer we can follow DDI
through the effective action equations and the larger the number of
reliable e-folds we shall gain.

This does answer, in my opinion, the objections raised recently  
\cite{TW} to the PBB
 scenario according to which it is
  fine-tuned. The situation here actually resembles that of
chaotic inflation \cite{chaotic}. Given some generic (though APT)  
initial data, we should ask
which is the distribution of sizes of the collapsing regions and of  
couplings therein.
Then, only the ``tails" of these distributions, i.e. those corresponding to
 sufficiently large,
and sufficiently weakly coupled, regions will produce Universes like  
ours, the rest will not.
The question of how likely a ``good" big bang is to take place is not  
very well posed
and can be greatly affected by anthropic considerations \cite{BDV}.

In conclusion, we may summarize recent progress on the problem of
initial conditions by saying that \cite{BDV}:

 \centerline {\bf Dilaton-Driven Inflation in String Cosmology}

\centerline {\bf is as generic as}

 \centerline {\bf Gravitational Collapse in General Relativity.}

Furthermore, asking for a sufficiently long period of
DDI amounts to setting upper limits on two arbitrary
moduli of the classical solutions.

Figure 7 (from Ref. \cite{BDV})
gives a $(2+1)$-dimensional sketch of a possible PBB Universe:  an  
original ``sea" of
dilatonic and gravity waves leads to collapsing regions of different  
initial size, possibly
to a scale-invariant distribution of them.
 Each one of these collapses is reinterpreted, in the string frame, as  
the process by which
 a baby Universe is born after a period of
PBB inflationary ``pregnancy",  the size of each baby Universe being  
determined
 by the duration of the corresponding
 pregnancy, i.e. by the initial size of (and coupling in) the
corresponding collapsing region.  Regions initially
larger than $10^{-13}~ {\rm cm}$ can generate Universes like ours,
smaller ones cannot.

\begin{figure}
\hglue3.0cm
\epsfig{figure=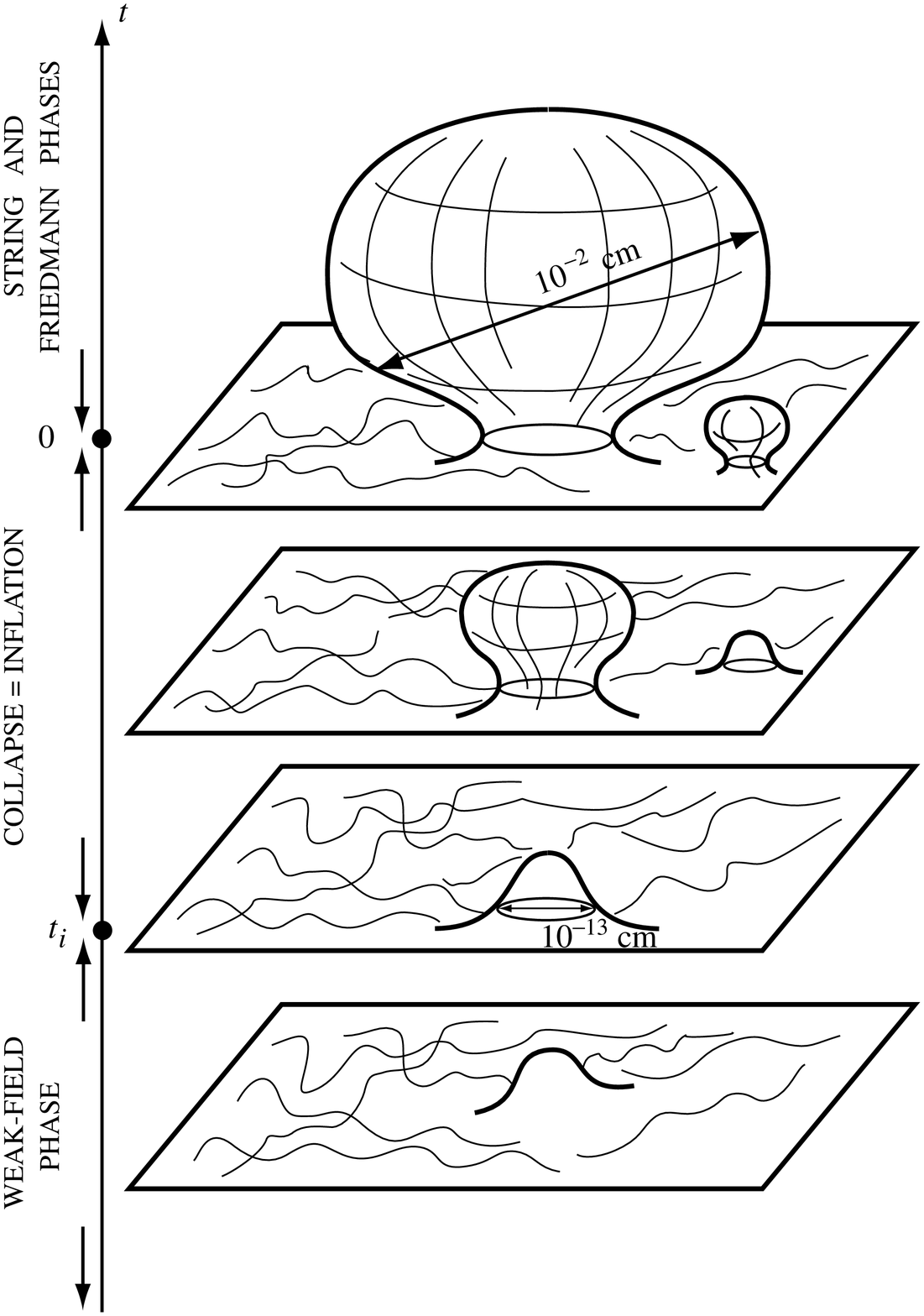,width=10cm}
\begin{center}
Figure 7
\end{center}
\end{figure}

A basic difference between the large numbers needed in (non- inflationary) FRW cosmology and the large numbers needed in  
PBB cosmology should be stressed.
In the former, the ratio of two classical scales, e.g. of total  
curvature to its
spatial component, which is expected to be $O(1)$,
has to be taken as large as $10^{60}$. In the latter, the above ratio is 
initially $O(1)$ in the
collapsing/inflating region,
and ends up being very large in that same region, thanks to DDI.
However, the (common) order of magnitude of these two classical quantities 
is a free parameter, and it is taken to be
much larger than the classically irrelevant quantum scale.
Indeed, the smallness of quantum corrections (which would introduce a  
scale in the problem)
was explicitly checked in \cite{GPV}.

We can visualize analogies and differences between standard and  
pre-big bang inflation
by looking again at Figs. 6a and 6b. The common feature
in the two pictures is that the fixed comoving scale corresponding
to the present horizon was
``inside the horizon" for some time during inflation,  possibly very  
deeply inside
at its onset.  The difference between the two scenarios
is just in the behaviour of the Hubble radius during inflation:  
increasing in standard
inflation (a), decreasing in string cosmology (b). Thus, while  
standard inflation is still facing the
 initial-singularity question and needs a non-adiabatic phenomenon to  
reheat the
Universe (a kind of small bang), PBB cosmology faces the singularity  
problem later,
combining it with the exit and heating problems (see Section 5).

\subsection{The spherically symmetric case}

In the spherically symmetric case many authors have studied the  
problem of gravitational collapse
of a minimally coupled scalar field both numerically and analytically.  
In the former case
I will only mention the well-known results of Choptuick \cite{Chp},
 pointing at  mysterious universalities
near critical collapse (i.e. at the  border-line situation in which  
the collapse criteria are
just barely met). In this case, a very small black hole forms. This is  
not the case
we are really interested
in for the reasons we just explained.
We shall thus turn, instead, to what happens when the collapse criteria are
largely fulfilled. For this we make use of the rather powerful results  
due to Christodoulou over
a decade of beautiful work \cite{Chr}, \cite{Chr87}, \cite{chi91},  
\cite{chi93}.

There are no gravitational waves in the spherically symmetric case
so that null wave data consist  of  just an
angle-independent asymptotic dilatonic wave form $f(v)$, with the associated
scalar News $N(v) =  f'(v)$.

A convenient system of coordinates is the double null
system, $(u,v)$, such that
\beq
\phi = \phi (u,v) \, ,
\label{eqn3.2}
\eeq
\beq
ds^2 = - \Omega^2(u,v)\,du \, dv + r^2(u,v)\,d \omega^2 \,,
\label{eq1.6}
\eeq
where $d\omega^2 = d\theta^2 + \sin^2 \theta \, d\varphi^2$. The field 
equations are conveniently re-expressed in terms of the three  
functions $\phi
(u,v)$, $r(u,v)$ and $m(u,v)$, where the local mass function $m(u,v)$ is 
defined by:
\beq
1 - \frac{2 m}{r} \equiv g^{\mu \nu}\,(\pa_\mu r)\,(\pa_\nu r) =
-\frac{4}{\Omega^2}\, \frac{\pa r}{\pa u}\,
\frac{\pa r}{\pa v}\,.
\label{eq1.7}
\eeq
One gets the following  set of evolution equations for $m$, $r$ and $\phi$ 
\bea
\label{eq1.8}
&& 2\frac{\pa r}{\pa u}
\,\frac{\pa m}{\pa u} = \left ( 1 - \frac{2 m}{r} \right )
\,\frac{r^2}{4}\,\left (\frac{\partial \phi}{\pa u} \right )^2 \,, \\
\label{eq1.9}
&& 2 \frac{\pa r}{\pa v}\,
\frac{\pa m}{\pa v} = \left ( 1 - \frac{2 m}{r} \right )
\,\frac{r^2}{4}\,\left (\frac{\partial \phi}{\pa v} \right )^2 \,,\\
\label{eq1.10}
&& r\,\frac{\pa^2 r}{\pa u \pa v} =
\frac{2 m}{r-2m}\,\frac{\pa r}{\pa u}\,
\frac{\pa r}{\pa v}\,,\\
\label{eq1.11}
&& r\,\frac{\pa^2 \phi}{\pa u \pa v} +
 \frac{\pa r}{\pa u} ~ \frac{\pa \phi}{\pa v}
+ \frac{\pa r}{\pa v} ~ \frac{\pa \phi}{\pa u}
=0\,.
\eea

The quantity
\beq
\mu (u,v) \equiv \frac{2 m (u,v)}{r}
\label{eqn3.3}
\eeq
plays a crucial r\^ole in the problem.
 If $\mu$ stays everywhere below 1, the field configuration will
not collapse but will finally disperse at infinity as outgoing waves. By 
contrast, if the mass ratio $\mu$ can reach anywhere the value 1, this  
signals
the formation of an apparent horizon ${\cal A}$. The location of this  
apparent
horizon is indeed defined by the equation
\beq
{\cal A}: \quad \mu (u,v) = 1 \, .
\label{eqn3.4}
\eeq
The above statements are substantiated by some rigorous inequalities  
\cite{chi93}
stating that:
\bea
\label{rel1}
&& \frac{\pa r}{\pa u} < 0\,, \quad \quad \quad \quad
\quad \frac{\pa m}{\pa v} > 0\,, \\
\label{rel2}
&& \frac{\pa r}{\pa v}\,
\left ( 1- \mu \right ) > 0\,, \quad \quad
\frac{\pa m}{\pa u}\,
\left ( 1- \mu \right ) < 0\,.
\eea
Thus, in weak-field regions ($\mu < 1$), $\partial_v
r > 0$, while, as $\mu > 1$, $\partial_v r < 0$, meaning that the
outgoing radial null rays (``photons'') emitted by the sphere $r =  
{\rm const}$
become convergent, instead of having their usual behaviour.
This is nothing else but the signature of a CTS!

 In the case of  spherical
symmetry, it has been possible to prove
\cite{chi91} that the
presence of trapped surfaces implies the existence of a future
singular boundary ${\cal B}$ of space-time where a curvature  
singularity occurs.
Furthermore, the behaviour of various fields near the singularity is  
just that of a
 quasi-homogeneous
DDI as described by Eqs. (\ref{inhKasner})! This highly non-trivial result
strongly supports the idea that
PBB inflation in the string frame
is the counterpart of gravitational collapse in the Einstein frame.

 Reference~\cite{chi91} gives the following sufficient criterion on  
the strength of
characteristic data, considered at some finite retarded time $u$
\beq
\frac{2 \Delta m}{\Delta r }  \geq
\left [ \frac{r_1}{r_2}\,\log \left (\frac{r_1}{2 \Delta r} \right )
+ \frac{6r_1}{r_2} -1 \right ]\,,
\label{eq2.2}
\eeq
where $r_1 \leq r_2$, with $r_2 \leq 3r_1/2$, are two spheres,
$\Delta r = r_2 - r_1$ is the width of the ``annular'' region between  
the two
spheres, and $\Delta m = m_2 - m_1 \equiv m(u,r_2) - m (u,r_1)$ is the mass 
``contained'' between the two spheres, i.e. more precisely the energy flux 
through the outgoing null cone $u =$ const, between $r_1$ and $r_2$.
Note the absence of any intrinsic scale (in particular of any  
short-distance cut-off)
in the above criterion.
The theorem proved in \cite{chi91} is not exhausted in the above
statement. It contains various bounds as well, e.g.
\begin{itemize}
\item an upper bound on the retarded time at which the CTS
(i.e. a horizon) is formed,
\item a lower bound on the mass, i.e. on the radius of the collapsing region.
\end{itemize}
 The latter quantity is very important for the discussion of the  
previous subsection
since it gives, in the equivalent string-frame problem, an upper limit
on the Hubble parameter at the beginning of DDI. Such an upper limit
depends only on the size of the advanced-time interval satisfying the CC;  
since the latter is determined by the scale-invariant condition  
(\ref{eq2.2}),
the initial scale of inflation will be classically undetermined.

The above criterion is rigorous but probably too conservative. It also  
has the shortcoming
that it cannot be used directly on ${\cal I}^-$, since $u \rightarrow  
-\infty$ on ${\cal I}^-$.
In Ref. \cite{BDV} a less rigorous (or less general) but simpler criterion
directly expressible in terms of the News (i.e. on ${\cal I}^-$) was  
proposed on
the basis of a perturbative study. It has the following attractive form:
\beq
 \sup_{v_1,v_2 \atop v_1 \leq v_2} \, {\rm Var} (N(x))_{x \in  
[v_1,v_2]} > C = O(1/4) \, ,
\label{eqn4.16}
\eeq
where:
\beq
{\rm Var} \, (N(x))_{x \in [v_1,v_2]} \equiv  \langle (N(x) -
\langle N \rangle_{[v_1,v_2]})^2 \rangle_{x \in [v_1,v_2]} \, .
\label{eqn4.15}
\eeq
Thus ${\rm Var} \, (g)_{[v_1,v_2]}$ denotes the ``variance'' of the function 
$g(x)$ over the interval $[v_1,v_2]$, i.e. the average squared  
deviation from its
mean value.

According to this criterion the largest interval satisfying (\ref{eqn4.16})
determines the size of the collapsing region and thus, through the
collapse inflation connection, the initial value of the Hubble parameter.
It would be interesting to confirm the validity of the above criterion  
and to determine
more precisely the value of the constant appearing on its r.h.s. through more
 analytic or numerical
work. Actually,  numerical studies of spherically symmetric
PBB cosmologies have already appeared \cite{MOV}, while more powerful  
numerical codes should  soon be available \cite{codes}.

\vskip 5mm

\section{Phenomenological Consequences}

\subsection{Cosmological amplification of vacuum fluctuations: general  
properties}

I will start by recalling the basic
physical mechanism underlying particle production in cosmology (for a  
nice review, see
\cite{quantum}) and
by introducing
the corresponding (and by now standard) jargon.
By the very definition of inflation ($\ddot a >0$) physical wavelengths
are stretched
past the Hubble scale ($H^{-1}$) during inflation. After the end of inflation
each wavelength grows slower than $H^{-1}$ and thus ``re-enters" the horizon.
Obviously, the larger the scale the earlier it crosses the horizon outward
and the later it crosses it back inward. Hence larger scales ``spend"  
more time
``outside the horizon" than smaller ones.

The attentive reader may worry at this point about
the way this description applies when distances are measured using the
Einsten-frame metric. As we have seen in the previous section,
PBB inflation corresponds to  accelerated contraction in the Einstein
frame.
Nonetheless, one can show that physical quantities (that is, typically, 
dimensionless ratios of physical quantities) do not depend on the choice
of the frame: after all, changing frame is nothing more than a local
field-redefinition, which is known not to affect the physics.
It is amusing to notice, for instance, that physical wavelengths go  
outside the
horizon during the Einstein-frame equivalent of DDI. Indeed, although
physical EF scales shrink during the collapse, the horizon $H^{-1}$
shrinks even faster! I refer to the first paper in \cite{MG2} for  
further discussion
on this  point.

Consider now a generic perturbation $\Psi$ on top of a homogeneous
background, which includes a cosmological-type metric, a dilaton, and,  
possibly,
other fields, such as another inflaton field, an axion, etc.
Since  $\Psi=0$ is, by definition of a perturbation, a classical solution,
$\Psi$ intself enters
the effective low-energy action quadratically. Soon after the  
beginning of inflation
the background itself becomes homogeneous, isotropic, and spatially flat,
 so that the perturbed action takes the generic form:
\begin{equation}
I ={\small\frac{1}{2}} \int d\eta\ d^3x\ S(\eta) \left[ \Psi
^{\prime 2}- (\nabla \Psi )^2\right].
\label{spertact}
\end{equation}
Here $\eta$ is  conformal-time ($a d \eta = dt$), and a prime denotes
 $\partial/\partial\eta$. The function $S(\eta)$ (sometimes called
the ``pump" field)  is, for any given $\Psi$, a given function of the
scale factor $a(\eta)$, and of other scalar fields (four-dimensional
dilaton $\phi(\eta)$,
moduli $b_i(\eta)$, etc.), which may appear non-trivially in the
background.

While it is clear that a constant $S$ may be reabsorbed by  rescaling   
$\Psi$,
and is  thus ineffective, a time-dependent $S$ couples non-trivially  
to $\Psi$
and leads to the production of pairs of quanta (with equal and  
opposite momenta).
In order to see this, it is  useful to go over to a Hamiltonian description
of the perturbation and of its canonically conjugate momentum $\Pi$:
\begin{equation}
 \Pi = \frac {\delta I}{\delta \Psi^{'}} = S~ \Psi'\; .
\label{defmomentum}
\end{equation}
The  Hamiltonian  corresponding to (\ref{spertact})
is thus given by
\begin{equation}
 H=\frac{1}{2} \int  d^3 x \Biggl[  S^{-1} \Pi^2 +
S (\nabla\Psi)^2\Biggr],
\label{ham}
\end{equation}
and the first-order Hamilton equations  read
\begin{equation}
\Psi' = \frac{\delta H}{\delta \Pi}=  S^{-1} \Pi \; , ~~~~~~~~~~~
\Pi' = -\frac{\delta H}{\delta \Psi}= S \nabla^2\Psi \;,
\label{frstord}
\end{equation}
leading to the decoupled second order equations
\begin{equation}
\Psi''+\frac{S'}{S} \Psi'- \nabla^2\Psi = 0 \, , ~~~~~~~~~~~
\Pi''-\frac{S'}{S} \Pi'- \nabla^2\Pi = 0 \, .
\label{psieq}
\end{equation}
 In Fourier space the Hamiltonian (\ref{ham}) is given by
\begin{equation}
H=\frac{1}{2} \sum_{\vec{k}}
  \Biggl(  S^{-1} \Pi_{\vec{k}}\Pi_{-\vec{k}} +
S k^2 \Psi_{\vec{k}}\Psi_{-\vec{k}}\Biggr),
\label{fham}
\end{equation}
where $\Psi_{-\vec{k}} = \Psi_{\vec{k}}^*$ and $\Pi_{-\vec{k}} =  
\Pi_{\vec{k}}^*$.
The equations of motion become
\begin{equation}
\Psi_{\vec{k}}' = S^{-1}{\Pi}_{-\vec{k}} \; , ~~~~~~~~~~~
\Pi_{\vec{k}}'= -S k^2 \Psi_{-\vec{k}} \; ,
\label{ffrstord}
\end{equation}
where $k=|\vec{k}|$. The transformation
\begin{equation}
\Pi_{\vec{k}} \rightarrow \widetilde \Pi_{\vec{k}}=k \Psi_{\vec{k}} \, ,
 ~~~~~ \Psi_{\vec{k}} \rightarrow \widetilde{\Psi}_{\vec{k}}=- k ^{-1}  
\Pi_{\vec{k}}\, , ~~~~~
S \rightarrow \widetilde S=S^{-1}
\label{sduality}
\end{equation}
leaves the Hamiltonian, Poisson brackets,
and equations of motion unchanged.
This symmetry of linear perturbation theory, and its physical consequences,
was discussed in \cite{sduality} under the name of S-duality,
 since it contains the usual
strong--weak  coupling (electric--magnetic) duality in the special case
of gauge perturbations.

In order to solve the perturbation equations, and to normalize the
spectrum, it is convenient to introduce the normalized (but no longer  
canonically conjugate)
 variables
${\widehat\Psi}$, ${\widehat\Pi}$,  whose  Fourier modes are defined by
\begin{equation}
\widehat\Psi_k = {S}^{1/2}\ \Psi_k \, , ~~~~~~~~~~~
\widehat\Pi_k = {S}^{-1/2}\ \Pi_k \, ,
\label{nv}
\end{equation}
so that the Hamiltonian density takes the canonical form:
\begin{equation}
H=\frac{1}{2} \sum_{\vec{k}}
 \left( |{\widehat\Pi}_k|^2 + k^2 |{\widehat\Psi}_k|^2\right) .
\label{313}
\end{equation}
Under S-duality, these new variables
 transform  as  the original ones. They
  satisfy the Schr\"odinger-like equations
\begin{equation}
{\widehat\Psi}_k{''}+\left[k^2-(S^{1/2}){''}
S^{-1/2}\right]{\widehat\Psi}_k
= 0 \, ,  ~~~~~
{\widehat\Pi}_k{''}+ \left[k^2-(S^{-1/2}){''} S^{1/2}\right]
{\widehat\Pi}_k = 0.
\label{nscndord}
\end{equation}

The amplification of perturbations is typically associated with a
transition from an inflationary phase in which the pump field is
accelerated to a post-inflationary phase in which
the pump field is decelerated or constant. In such a class of
backgrounds, the ``effective potentials",
$V_{\Psi}=(S^{1/2}){''}S^{-1/2}$ and $V_{\Pi}=(S^{-1/2}){''} S^{1/2}$,
grow  during the phase of accelerated evolution, and
decrease in the post-inflationary, decelerated epoch, vanishing
asymptotically both for very early times, $\eta \rightarrow -\infty$,
and for
 very late times, $\eta \rightarrow + \infty$.

The initial evolution of perturbations, for all modes with
$k^2> |V_{\Psi}|$, $|V_{\Pi}|$, may be described by the
WKB-like approximate solutions of Eqs. (\ref{nscndord})
\begin{eqnarray}
{\widehat\Psi}_k(\eta)&=&\left( k^2-V_{\Psi}\right)^{-1/4}\
e^{\ -i\int\limits_{\eta_0}^{\eta} d\eta' \left( k^2-V_{\Psi}\right)^
{1/2} },  \nonumber \\
{\widehat\Pi}_k(\eta)&=& k \left( k^2-V_{\Pi}\right)^{-1/4}\
e^{\ -i\int\limits_{\eta_0}^{\eta} d\eta' \left( k^2-V_{\Pi}\right)
^{1/2} },
\label{wkbsol}
\end{eqnarray}
which we have normalized to a vacuum fluctuation, and
where the extra factor of $k$ in the solution for ${\widehat\Pi}_k$
comes  from consistency with the first order equations (\ref{ffrstord}).
We  have ignored
a possible relative phase in the solutions. Solutions (\ref{wkbsol})  
manifestly
preserve the S-duality symmetry of the equations, since the
potentials  $V_{\Psi}$, $V_{\Pi}$ get interchanged under $S \rightarrow
S^{-1}$.

Let us now discuss two opposite regimes:
\begin{itemize}
\item When the perturbation is deeply inside the horizon ($ k/a \gg H$)
we find ``adiabatic" behaviour, i.e.
\begin{equation}
 k \Phi_k \sim S^{-1/2}\; , \; \Pi_k \sim S^{1/2} \; ,
\label{adiabatic}
\end{equation}
 implying, through (\ref{ham}), that the contribution to the
Hamiltonian of modes inside the horizon stays constant.
\item When the perturbation is far outside the horizon ($ k/a \ll H$),  
it enters
the so-called freeze-out regime in which $\Psi$ and $\Pi$ stay  
constant (better
have a constant solution, see \cite{sduality}). Such a behaviour  
implies, again through
(\ref{ham}), that the contribution of super-horizon modes to the Hamiltonian
grows in time. If $\dot{S} >0$,  the growth of $\cal{H}$ is  due to  
$\Psi$, while,
for  $\dot{S} < 0$, it is due to $\Pi$. In either case the growth is due to
particle production in squeezed states \cite{Grish1}, i.e. states in which
one canonical variable is very sharply defined and the conjugate one
is largely undetermined. Although, strictly speaking, quantum coherence
is not lost, in practice the sub-fluctuating variable cannot be
measured with unlimited precision (coarse graining) and therefore
entropy is produced (see Subsection 4.6).
\end{itemize}
It is not too hard to join the two extreme regimes mentioned above
and to find the qualitative and quantitative features of the solutions.
For lack of space we refer the reader to the original literature (see,  
e.g. \cite{sduality}).

The above considerations were very general. What is instead typical
 of the PBB scenario?  There are at least two
features that are quite unique to string cosmology:
\begin{itemize}
\item Pump fields, and in particular their contributions to the
evolution equations (\ref{nscndord}), grow during PBB inflation, while  
they tend to decay
in standard inflation.
\item The richer set of backgrounds and fluctuation
present in string theory allows for the amplification of new kinds of
perturbations.
\end{itemize}

One can
easily determine the pump fields for each one of the most interesting  
perturbations
appearing in the PBB scenario.
The result is:
\begin{eqnarray}
\rm{Gravity~waves,~dilaton}&:&   S = a^2 e^{-\phi} \nonumber \\
\rm{Heterotic~gauge~bosons}&:&  S =  e^{-\phi} \nonumber \\
\rm{Kalb--Ramond,~axions}&:&   S = a^{-2} e^{-\phi}\; .
\label{pump}
\end{eqnarray}

In the following subsections we shall briefly describe
the characteristics of these four perturbations after their
original vacuum fluctuations are amplified by PBB inflation.
For further details, see also \cite{GV95}.

\subsection{Tensor perturbations: an observable cosmic gravitational  
radiation background (CGRB)?}

It is not surpising to find that, for tensor and dilaton perturbations, 
the pump field is nothing but the scale factor in the Einstein frame  
($a_E = a e^{-\phi/2}$)
since, in this frame,
the action for gravity and for the dilaton take the canonical form.  
The Einstein-frame
scale factor corresponds to a collapsing Universe (see Section 3), hence to
the decreasing pump field $a_E(\eta) \sim \eta^{1/2}$ during DDI.
For scales that go outside the horizon during DDI, this implies \cite{BGGV}
 a Raileigh--Jeans-like spectrum,
$d\Omega/d {\rm log} k \sim k^3$, up to logarithmic corrections \cite{BGGV}.

When the curvature scale reaches the string scale we expect DDI to  
end, and a high
(string scale)  curvature phase to follow, before the eventual exit to
the FRW phase takes place (see Section 5). Not much is known about the  
string phase, but, using
some physical arguments as well as some quantitative estimates, it can  
be argued that
such a phase will lead to copious GW production at frequencies  
corresponding to
the string scale at the time of exit. After the transition to the FRW phase,
all particle production switches off. This is why our GW spectrum has  
an end point that
 corresponds to
  the string/Planck scale at the beginning of the FRW phase. If no  
inflation takes place
after, the end-point frequency corresponds, today, to $\omega = \omega_1 \sim 
100$ GHz.

As illustrated in Fig. 8, the GW spectrum can be rather flat below the  
end point,
up to the frequency $\omega_s$, the last scale that went out of
the horizon during DDI. Further below $\omega_s$ we get the above-mentioned
steep $\omega^3$ spectrum. It thus looks as if the best chances for  
the detection of our
stochastic background lie precisely near $\omega_s$, where a kink (or  
knee) is expected.

\begin{figure}
\hglue3.0cm
\epsfig{figure=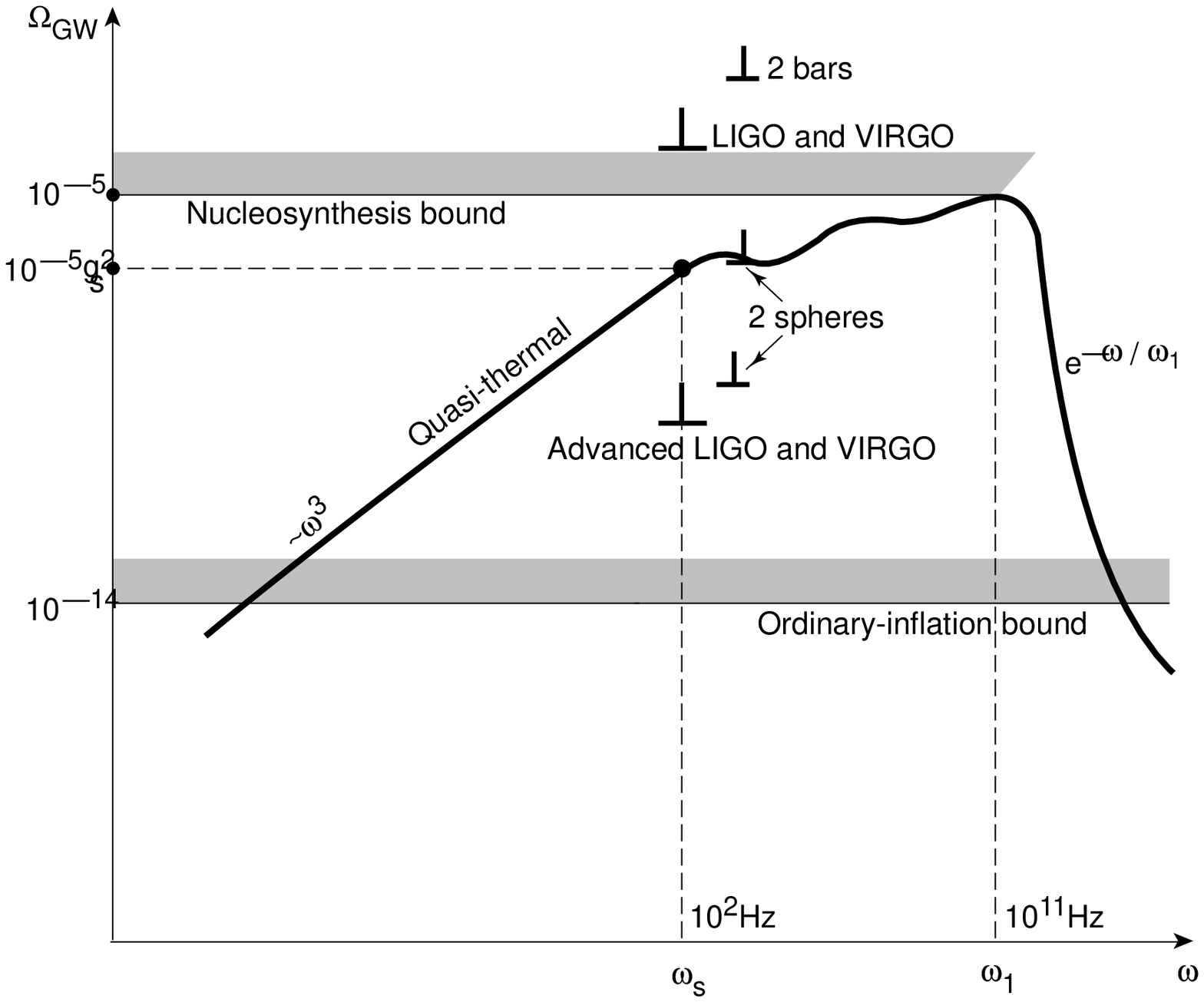,width=10cm}
\begin{center}
Figure 8
\end{center}
\end{figure}

Unfortunately, the position of the knee and the value of $\Omega_{GW}$  
at that point
depend on two background parameters that are, so far, difficult to predict.
One corresponds to the duration of (better, the total red-shift  
during) the string phase,
 the other to the value of $l_P/\lambda_s$ at the end of DDI (hence to  
the value of the dilaton
at that time).
As shown in Fig. 8,
values of $\Omega_{GW}$ in
the range of
$10^{-6}$--$10^{-7}$ are possible in some regions of parameter space,  
which, according to
some estimates of sensitivities \cite{sens} reported in the same figure for
$\omega_s \sim 10^2$Hz, could be inside detection
capabilities in the near future.


The signal is predicted to consist of randomly distributed {\it  
standing} waves, a feature
that has been argued \cite{Grish2} to further help detection. In any case,
 cross-correlation experiments
are mandatory here in order to disentangle this stochastic signal from  
real noise.
Sensitivities to a CGRB of this type have been estimated for a variety
of two-detector combinations \cite{sens}.
A comprehensive review of GW experiments and of their relevance to
the early Universe can be found in \cite{Maggiore}.

\subsection{Dilaton perturbations}

Since the dilaton is, after all, the inflaton of PBB cosmology, its  
fluctuations
are the most natural source of adiabatic scalar perturbations.
We recall that, in standard cosmology, inflaton fluctuations naturally lead
to a quasi scale-invariant, Harrison-Zeldovich (HZ) spectrum of  
adiabatic perturbations, something
highly desirable both to explain CMB anisotropy and for models of LSS
formation. Can we get something similar from the dilaton? The answer,  
unfortunately,
is no! Let me spend a moment explaining why.

Unlike tensor perturbations, which do not couple to the scalar field  
to linear order
and are gauge-invariant by themselves, scalar perturbations are  
contained, a priori,
in five functions defined by:
\bea
ds^2 &=& a^2(\eta) \left[ - (1 + 2 \Phi) d\eta^2 + \left((1 - 2 \Psi)  
\delta_{ij} +
\partial_i\partial_j E \right) dx^idx^j - 2 \partial_i B dx^i d \eta  
\right] \nonumber \\
\phi &=& \phi_0(\eta) + \chi(\eta, \vec{x})\; .
\label{scalarpert}
\eea
The five functions $\Phi, \Psi, B, E, \chi$ are not separately
gauge-invariant. However,  the following ``Bardeen"
combinations are gauge-invariant:
\bea
\Phi_B &=& \Phi + \frac{1}{a} \left[a(B - E') \right]^{'} ~~, \nonumber \\
 ~\Psi_B &=& \Psi - \frac{a'}{a} (B - E')~~, \nonumber \\
~ \chi_{GI} &=& \chi + \frac {\phi_0'~a} {a'} \Psi \; .
\label{Bardeen}
\eea
Introducing the variable $v \equiv a\chi_{GI}$, the scalar field
enters the quadratic action ``canonically" i.e.:
\begin{equation}
S_{eff}(v) = \frac{1}{2} \int d\eta d^3 x \left[v'^2 - (\vec{\nabla}  
v)^2 + (z^{''}/z) ~ v^2 \right]
  \; ,~~  z \equiv \frac{\phi'_0 a^2}{a'}\; ,
\label{vact}
\end{equation}
giving the evolution equation
\begin{equation}
v_k^{''} + \left(k^2 - (z^{''}/z)\right)~v_k =0 \; .
\label{veq}
\end{equation}

In the DDI background, $z \sim a$ and thus the canonical scalar field
obeys the same equation as the canonical graviton field, therefore  giving
identical spectra (as far as the dilaton remains massless, of course).
This strongly suggests that adiabatic perturbations in PBB cosmology
have a Raleigh--Jeans, rather than HZ, spectrum and that they are unsuitable
for generating CMBA or LSS.
Before being sure of that, however, we have to analyse
the scalar fluctuations of the metric itself in terms of the above-mentioned
Bardeen potentials $\Phi_B, \Psi_B$.

A popular gauge (particularly advertised in \cite{quantum}) is the so-called
longitudinal gauge, defined by $B = E = 0$, where $\Phi_B = \Phi$ and  
$\Psi_B = \Psi$.
In this gauge one of the constraints simply reads $\Phi = \Psi$, while  
a second constraint
relates either one of them to $v$:
\begin{equation}
 \Psi_k = - \frac {\phi_0'}{4 k^2} (v_k/a)'
\sim k^{-3/2}~\frac {1}{|k \eta|^2} (k/a)_{HC}\; ,
\label{Psik}
\end{equation}
where we have inserted the small $k$ behaviour of $v_k$, which is  
identical to
that of tensor perturbations.

Unfortunately, Eq. (\ref{Psik}) leads to very large fluctuations of  
$\Psi = \Phi$
at small $k \eta$, so large that one leaves the linear-perturbation regime
for the expansion (\ref{scalarpert}) of the metric much before the  
high-curvature
scale is reached. Does this mean that the metric becomes very inhomogeneous?
It would look to be the case \dots$~$ unless the growth of $\Psi$ and  
$\Phi$ is
in some way a gauge artefact. But how can it be a gauge artefact if
$\Psi$ and $\Phi$ correspond, in this gauge, to the gauge-invariant  
Bardeen potentials?
The answer to this question was provided in \cite{BGGMV}. By going from
the longitudinal gauge to an ``off-diagonal" gauge with $\Psi= E =0$,  
or, even better,
to one in which only $\Phi$ and $E$ appear, one finds that  
perturbations of the
metric remain small at all $\eta$ till Planckian/string-scale curvatures
are reached.

This is easy to see, for instance, in a gauge with $\Psi = B =0$, where 
$\Psi_B \sim (a'/a) E'$. Clearly this gives $E \sim \eta^2 \Psi_B$ and, since
$E$ enters the metric with two spatial derivatives, this implies
that $h_{ij} \sim (k\eta)^2 \Psi_B$, which is sufficiently small at  
small $k \eta$
for linear perturbation theory to be valid.
One can then look for physical effects of these scalar perturbations
(e.g. for contributions to CMBA) and find
that they actually remain as small as the tensor contributions.
In conclusion, once gauge artefacts are removed, it seems that
adiabatic scalar perturbations, as well as their tensor counterparts,
remain exceedingly small at large scales.

On the other hand, the rather large yields at short scales also apply to
dilatons. This allows for a possible source of scalar waves if the dilaton
is very light. However, as recently discussed by Gasperini \cite{Gasp},
it is very unlikely that such a signal will be observable, given the
constraints on the dilaton mass due to tests of the equivalence principle
(see Section 2).
Other restrictions on the dilaton mass come from the possibility
that their
density may become overcritical and close the Universe. This and other
possible interesting windows in parameter space are discussed
in \cite{MG3}, and will not be reported in any detail here.

\subsection{Gauge-field perturbations: seeds for $\vec{B}_{gal}$?}

In standard inflationary
cosmology there is no amplification of the vacuum fluctuations of  
gauge fields.
This is a straightforward consequence of the fact that
inflation makes the metric conformally flat, and of the decoupling
of gauge fields from a conformally flat metric precisely in $D=3+1$  
dimensions.

As a very general remark, apart from pathological
solutions, the only background field that can amplify, through its  
cosmological variation,
 e.m. (more generally gauge-field) quantum fluctuations
 is the effective gauge coupling
itself \cite{Ratra}. By its very nature,
 in the pre-big bang scenario  the effective gauge coupling inflates
together with space during the PBB phase.
It is thus automatic that any efficient PBB inflation brings together
 a huge variation of the effective gauge coupling, and
thus a very large amplification of the primordial
e.m. fluctuations \cite{GGV}. This can possibly provide the  
long-sought for origin
for the primordial seeds of the observed
galactic magnetic fields.

To be more quantitative, since the pump field for electromagnetic  
perturbations
 is  the effective (four-dimensional) gauge coupling itself (see Eq.  
(\ref{pump})),
the total amplification of e.m. perturbations on any given scale  
$\lambda$ is given by
$\alpha_0/\alpha_{ex}$, i.e. by the ratio of the fine structure constant now
and the fine structure constant at the time of exit of the scale
$\lambda$ during DDI.
It turns out \cite{GGV} that, in order to produce sufficiently large seeds
for the galactic magnetic fields, such a ratio has to be enormous
for the galactic scale, i.e. about $10^{66}$.
Taken at face value, this would be a very strong indication in favour of the 
PBB scenario, more particularly of DDI. Indeed, only in such a  
framework  is it natural
to expect that the effective gauge coupling grew during inflation by
a factor whose logarithm is of the same order as the number of
inflationary e-folds.

Notice, however, that, unlike GW, e.m. perturbations interact quite
considerably with the hot plasma of the early (post-big bang) Universe.
Thus, converting the primordial seeds into those that may have existed
at the protogalaxy formation epoch is by no means a trivial exercise
 (see, e.g. \cite{Ostreiker}).
The question of whether or not the primordial seeds generated
in PBB cosmology can evolve into the observed galactic magnetic fields thus
remains, to this date, an unsolved, yet very interesting, problem.

\subsection{Axion perturbations: seeds for CMBA and LSS?}

In four dimensions the curl of $B_{\mu\nu}$, $H_{\mu\nu\rho}$,
is equivalent to a pseudoscalar field, the (KR) axion
$\sigma$, through
 \begin{equation}
 H_{\mu\nu\rho} = e^{\phi}~ \epsilon_{\mu\nu\rho\tau}
\partial^{\tau} \sigma \; .
\label{KRA}
\end{equation}
It is easy to see that, while the pump field for $B_{\mu\nu}$ is
$a^{-2}~e^{-\phi}$, that for $\sigma$ is $a^{2}~e^{\phi}$. Indeed
their respective perturbations
are related by the duality of perturbations discussed in Subsection 4.1. 
We can use either description with identical physical results.
Note that, while $a$ and $\phi$ worked in opposite directions
for tensor and dilaton perturbations, generating strongly tilted (blue)
spectra, the two work  in the same direction for axions, so that
spectra can be flat or even tilted towards large scales (red spectra)  
\cite{Copeland}.
An interesting fact is that, unlike the GW spectrum, that of axions is  
very sensitive to
the cosmological behaviour of internal dimensions during the DDI epoch.
On one side, this makes  the model less predictive. On the other, it  
tells us that
axions represent a window over the multidimensional cosmology expected  
generically
from string theories, which must live in more that four dimensions.
 Parametrizing the spectrum
by:
\begin{equation}
\Omega_{ax}(k) = \left({H_{max}\over M_P}\right)^2 (k/ k_{max})^{\alpha} \; ,
\end{equation}
and considering the case of three non-compact and six compact dimensions
with separate isotropic evolution, one finds:
\begin{equation}
 \alpha =  {3 + 3 r^2 - 2 \sqrt{3 + 6 r^2} \over 1 + 3 r^2} ~,
\label{spectralindex}
\end{equation}
where
\begin{equation}
r \equiv {1\over 2} {\dot{V_6}~V_3 \over V_6~\dot{V_3}}
\label{r}
\end{equation}
 is a measure
of the relative evolution of the internal and external volumes.
Equations (\ref{spectralindex}), (\ref{r})
show that the axion spectrum becomes exactly HZ (i.e. scale-invariant) when
$r = 1$, i.e. when all  nine spatial dimensions of superstring theory  
evolve in a rather
symmetric way \cite{BMUV2}.
In situations near this particularly symmetric one, axions are able to  
provide
a new mechanism for generating large-scale CMBA and LSS.

Calculation of the effect gives \cite{Durrer}, for massless axions:
\begin{equation}
l(l+1) C_l \sim O(1) \left({H_{max}\over M_P}\right)^4 (\eta_0  
k_{max})^{-2\alpha}
{\Gamma(l+\alpha) \over \Gamma(l- \alpha)}\; ,
\end{equation}
where  $C_l$ are the usual  coefficients of the multipole expansion of  
$\Delta T/T$
\begin{equation}
\langle \Delta T/T(\vec{n})~~ \Delta T/T(\vec{n}')\rangle ~ = ~
 \sum_l (2l+1) C_l P_l(\cos\theta)\; , ~~ \vec{n} \cdot \vec{n}' =  
\cos\theta\;,
\end{equation}
and $\eta_0 k_{max} \sim 10^{30}$.
In string theory, as repeatedly mentioned, we expect $H_{max}/ M_P  
\sim M_s/M_P \sim 1/10$, while the exponent $\alpha$ depends on the  
explicit
PBB background with the above-mentioned HZ case corresponding to  
$\alpha =0$. The standard
tilt parameter $n = n_s$ ($s$ for scalar) is given by $n = 1 + 2  
\alpha$ and is
found, by COBE\cite{COBE},
to lie between $0.9$ and $1.5$, corresponding to $0 < \alpha < 0.25$  
(a negative $\alpha$ leads
to some theoretical problems). With these inputs we can see that the  
correct normalization
($C_2 \sim 10^{-10}$) is reached for $\alpha \sim 0.2$, which is just in the 
middle of the allowed range. In other words, unlike in standard  
inflation, we cannot
predict the tilt, but when this is given, we can predict (again unlike  
in standard inflation)
the normalization.

With some extra work \cite{Melchiorri}  one can compute the $C_l$ in  
the acoustic-peak region
adding vector and tensor contributions from the seeds.
It turns out that the acoustic-peak structure is very sensitive to
$\alpha$, hence to the behaviour of the internal dimensions during the  
DDI phase.
The above-mentioned value, $\alpha = 1$, does not give peaks at all  
and, as such,
looks ruled out by the data.
Values of $\alpha$ in the range $0.3$--$0.4$ appear to be preferred  
(especially in the
presence of a cosmological constant with $\Omega_{\Lambda} \sim 0.7$).
We saw, however, that the overall normalization was very sensitive to the
value of $\alpha$. For $\alpha$ in the $0.3$--$0.4$ range, the  
normalization is off (way too small)
by many orders of magnitude.
Therefore, if present indications are confirmed, as they seem to be  
from the recent release
of the Boomerang 1997 data analysis \cite{Boomerang},
one will be forced to a $k$-dependent $\alpha$, meaning different phases
in the evolution of internal dimensions during  DDI.

\subsection{Heating up the Universe}

Before closing this section, I wish to recall how one sees the very  
origin of the hot big bang
in this scenario. One can easily estimate the total energy stored in the
quantum fluctuations, which were amplified by the pre-big bang backgrounds 
(for a discussion of generic perturbation spectra, see \cite{BMUV2, BH}. 
The result is, roughly,
\begin{equation}
\rho_{quantum} \sim N_{eff} ~ H^4_{max} \; ,
\label{rhoq}
\end{equation}
where $N_{eff}$ is the effective number of species that are amplified  
and $H_{max}$ is the maximal
curvature scale reached around $t=0$. We have already argued that  
$H_{max} \sim M_s =
 \lambda_s^{-1}$,
 and we know that,
in heterotic string theory, $N_{eff}$ is in the hundreds. Yet, this  
rather huge energy density
is very far from critical, as long as the dilaton is still in the  
weak-coupling region,
justifying our neglect of back-reaction effects. It is very tempting  
to assume \cite{BMUV2} that,
precisely when the dilaton reaches a value such that $\rho_{quantum}$  
is critical, the Universe
will enter the radiation-dominated phase. This PBBB (PBB bootstrap)  
constraint gives, typically:
\begin{equation}
e^{\phi_{exit}}  \sim 1/N_{eff}\;\; ,
\label{PBBB}
\end{equation}
i.e. a value for the dilaton close to its present value.

The entropy in these quantum fluctuations can also be estimated following
some general results \cite{entropy}. The result for the density of  
entropy $S$ is, as expected,
\begin{equation}
S \sim N_{eff} H_{max}^3\;.
\end{equation}
It is easy to check that, at the assumed time of exit given by (\ref{PBBB}), 
this entropy saturates
recently proposed holography bounds.
The discussion of such bounds is postponed to Subsection 5.7 since is
has also interesting implications for
the exit problem.
\vskip 5mm

\section{How could it have stopped?}

We have argued that, generically, DDI, when considered at lowest
order in derivatives and coupling, evolves {\it towards} a singularity of the
big bang type. Similarly, at the same level of approximation,
non-inflationary solutions of the FRW type
emerge {\it from} a singularity. Matching these two branches in a  
smooth, non-singular way
has become known as the (graceful) exit problem in string cosmology  
\cite{nogo}.
 It is, undoubtedly,
the most important theoretical problem the PBB scenario is facing today.

Of course, one would  not only like to know that a graceful exit does  
take place:
one would also like to describe the transition between the two phases
in a quantitative way.
Achieving this goal would amount to nothing less
than a full description of what replaces the
big bang of standard cosmology in the PBB scenario.
As mentioned in Section 1, this difficult problem is the analogue, in
string cosmology, of the (still not fully solved) confinement problem of QCD.
The exit problem is particularly hard because, by its very nature, and by the
existing no-go theorems \cite{nogo}, it must occur, if at all, at  
large curvature and/or coupling
and, because of fast time-dependence, must break (spontaneously)  
supersymmetry.
The phenomenological predictions made in the
previous section were based on the assumption that i) a graceful exit  
does take place;
ii) sufficiently large scales are only affected by it kinematically,  
i.e. through
an overall red-shift of all scales.

In this section, after recalling some no-go theorems for the exit,
 we will review various proposals that circumvent those theorems starting
from the mathematically simplest, but physically least realistic,  
proposals  and ending with the physically favoured, but harder to  
analyse, suggestions.
  The latter proposals
suggest possible lines along which a quantitative description of the  
exit might
eventually  emerge.
Needless to say, in spite of the many encouraging results, much work  
remains to be done: perhaps new techniques, and/or a deeper  
understanding of string theory
in its non-perturbative regimes through the construction
of the still largely unknown M-theory \cite{M}, need
to be developed before a full quantitative description
can be hoped for.

I should also mention that there have been suggestions \cite{anisexit}  
that the BB singularity
can be avoided if the DDI phase is highly anisotropic.
While this is an interesting suggestion, with isotropization taking place
later in the non-inflationary regime, we will stick here to
the simplest case, in which DDI has already prepared a very
 homogeneous  Universe before exit takes place. This is why
our discussion on the exit problem is limited to the case of
homogeneous cosmologies.
Also, for lack of space, I shall refer the reader to the literature  for
most of the  details.

\subsection{No-go theorems}

Under some restrictive conditions \cite{nogo}, it was shown that one  
cannot have
a change of branch, i.e. that the Universe cannot make a permanent transition
from the inflationary pre-big bang to a FRW post-big bang solution.
Perhaps the best way to convey the physical meaning of those
theorems is in terms of the
 necessary conditions for exit recently given by Brustein and Madden  
\cite{BM}.
These authors give necessary conditions for two subsequent events to occur:
firstly, a branch change in the string frame should take place: this  
imposes the violation of some
energy conditions; secondly, a bounce should occur in the E-frame  
metric since, as we have
seen in Section 3, DDI represents a collapse in the E-frame. This  
latter transition requires
further violation of energy conditions.

Before the reader gets too worried about these violations, I should  
point out that
these refer to the  equations of state satisfied by some ``effective"  
sources,
which include both higher-derivative and higher-loop corrections. It  
is well known that
such sources generically do lead to violations of the standard energy  
conditions satisfied by
normal matter or radiation-like classical sources.

\subsection{Exit via a non-local $V$}

This is perhaps the simplest example of an exit. It was first  
discussed in \cite{MV2}.
The reason why this is not considered an appealing mechanism for the exit
is that  the potential it employs  depends on
$\bar{\phi}$ (instead of $\phi$), in order to preserve SFD. By general  
covariance such a potential, if non-trivial, must
be non-local. Unfortunately, there has been no convincing proposal to  
explain how such non-local potentials
might arise within a superstring theory framework.

\subsection{Exit via $B_{ij}$}

The antisymmetric KR field may lead to violation of the energy conditions
and thus induce an exit.
Some amusing examples were given in \cite{GMV2}, where a non-trivial  
$B_{ij}$ is introduced
through $O(d,d)$ transformations acting on a pure metric-dilaton  
cosmology of the type
described in Subsection 2.5.
It was found that these so-called ``boosted" cosmologies were less singular
 than the original ones. In some cases they were even completely free  
of singularity
and provided examples of exit, albeit in not-so-realistic situations.

It is tempting to speculate that this softening of singularities, due  
to a non-trivial
$B_{ij}$ field, could be related to recent developments in the field  
of non-commutative
geometry \cite{noncomm} induced by a $B_{ij}$ field.
Work along these lines is in progress.

\subsection{Exit via quantum tunnelling}

Several groups \cite{WDW} have attempted to describe the transition
from the pre- to the post-big bang without modifying the
 low-energy tree-level effective action,
by exploiting the quantum cosmology approach based
on the Wheeler--De Witt (WDW) equation.
In Refs. \cite{WDW} an $O(d,d)$-invariant
WDW equation was derived in the $(d^2+1)$-dimensional mini-superspace  
consisting
of a homogeneous Bianchi I metric, the  antisymmetric tensor, and the  
dilaton. The $O(d,d)$
symmetry helps avoiding the ordering ambiguities which usually plague the WDW
equation. For the time being only the mathematically simpler case of an
$O(d,d)$-invariant potential $V(\bar{\phi})$ has been  analysed since,
in that case, $d^2$-conserved charges can be defined and the ``radial" part 
of the WDW equation reduces to a one-dimensional Schroedinger equation for
a scattering problem.

It is amusing that, from such a point of view, the initial state of  
the Universe
is described by a right-moving plane wave, which later encounters a  
potential,
giving rise to both a transmitted and a reflected (i.e. left-moving) wave.
The transmission coefficient gives the probability that the Universe ends up
in the pre-big-bang singularity, while the reflection coefficient  
gives the probability
of a successful exit into the post-big-bang decelerating expansion.

For certain forms of $V(\bar{\phi})$ the wave is classically reflected and
the WDW approach just confirms this expectation by giving a $100 \%$
probability for the exit.  However, even when there is no classical exit, the
probability of wave-reflection  is non-zero because of quantum tunnelling.
The quantum probability of a classically forbidden exit
turns out to be exponentially suppressed  in the coupling constant
$e^{\phi}$, which is just fine. Unfortunately, it is
also exponentially suppressed in the total volume of 3-space (in  
string units)
after the  pre-big-bang. Thus, only tiny regions of space have a reasonable
chance to tunnel.

\subsection{Higher-derivative corrections}

While the examples of exit given in the  previous subsections are
theoretically interesting, they do look somewhat artificial and non-generic.
In this and in the following subsection we shall describe
two mechanisms for exit that involve very general properties
of the lowest order solutions and of string theory.
The present feeling is that, if  graceful exit occurs, it should be
maily induced by some combination of higher-derivative and higher-loop
effects. Let us start with the former.

 Toy examples have shown \cite{GMV} that DDI can flow, thanks to
higher-curvature corrections, towards a de-Sitter-like phase, i.e.  
into a phase of
constant $H$ (curvature) and constant $\dot{\phi}$. This phase is expected to
last until loop corrections become important and give rise to
a transition to a radiation-dominated phase (see the next subsection). 
The idea is to justify the strong curvature transition
from the dilatonic to the string phase by proving the existence of an exact 
de Sitter-like solution to the field equation, which acts as a late time
attractor for the perturbative DDI branch.
 As shown in \cite{GMV}, the existence of such
attractors depends on the existence of (non-trivial) solutions for a  
system of $n$
algebraic equations in
$n$ unknowns. In general, we may expect a discrete number of solutions
to exist. If at least one of them has some qualitative  
characteristics, it will act as a
late-time attractor for solutions approaching DDI in the far past.
An explicit example of this phenomenon was constructed in \cite{GMV}.
In this connection, it is worth mentioning that solutions connecting  
duality-related
low-energy branches through a high-curvature CFT were already proposed  
in \cite{KK}.

It was recently pointed out \cite{MR} that the reverse order of events  
is also
possible. The coupling may become large {\it before} the curvature  
does. In this case, at least
for some time, the low-energy limit of M-theory should be adequate:  
this limit is known \cite{M}
to give $D=11$ supergravity and is therefore amenable to reliable  
study. It is likely,
though not yet clear, that, also in this case, strong curvatures will
have to be reached before the exit
can be completed.

\subsection{Loop corrections and back reaction}

The idea here is to invoke the back reaction from particle production as
the relevant mechanism. Since the back reaction is an $O(e^{\phi}  
\alpha' H^2)$
correction, its effect is contained in one-loop $O(R^2)$ contributions
to the effective action. A recent calculation \cite{Ghosh2} shows  
that, indeed,
loop corrections to DDI work in the right direction and become relevant 
precisely when expected according to the exit criterion (\ref{PBBB}).

A class of such contributions was analysed some time ago by Antoniadis  
et al. \cite{ART}
in the case of a spatially flat ($k=0$) cosmology and  by
Easther and Maeda \cite{Maeda} in the case of a closed Universe ($k=1$).
Both groups find non-singular solutions to the loop-corrected field  
equations.
However, neither group is actually able to obtain  solutions that start in
the dilaton-driven superinflationary regime and later evolve
 through a branch change.

More recently, several examples of full exit have been constructed  
\cite{fullexit}.
Although they are based on $\alpha'$ and loop-corrected actions, which  
have not been
derived from reliable calculations, they seem to indicate, at least,  
that the BM
 conditions for exit
may turn out to be not just necessary but also sufficient.
It also appears \cite{MPC} that exit occurs when the entropy bound  
becomes threatened by the entropy in the
amplified/squeezed quantum fluctuations, as we shall now discuss.

\subsection{Entropy considerations}

Entropy-related considerations have recently led to model-independent
arguments in favour of the occurrence of a graceful exit in string cosmology.
As we shall see, those are physically quite close to the arguments based on
back-reaction and loop corrections, which we have just discussed in  
the previous subsection.

Almost twenty years ago Bekenstein \cite{Bek1}
suggested that, for a  limited gravity system of energy $E$, and
whose size $R$ is larger than its gravitational radius, $R > R_g
\equiv 2G_N E$,
  entropy is bounded by $S_{BEB}$:
\begin{equation}
 S_{BEB} = ER/\hbar = R_g~ R ~l_P^{-2}.
\label{BEB}
\end{equation}

Holography \cite{holography}  suggests that maximal entropy is
bounded by $S_{HOL}$,
\begin{equation}
S_{HOL}= A l_P^{-2}, \label{HOL}
\end{equation}
where $A$ is the area of the space-like
surface enclosing the region of space whose entropy we wish to bound.
For systems of limited gravity, since $R > R_g, A=R^2$,
 (\ref{BEB}) implies the holography bound (\ref{HOL}).

Can  these entropy bounds be applied to the whole Universe, i.e. to  
cosmology?
A cosmological Universe is {\it not} a system of limited
gravity, since its large-distance behaviour is determined by the  
gravitational
effect of its matter content through Friedmann's equation (\ref{FRE}).
Furthermore, the holography bound obviously fails for sufficiently large
regions of space since, for a given temperature, entropy grows like $R^3$
while area grows like $R^2$.
The generalization of entropy bounds to cosmology turned out to be subtle.

In 1989, Bekenstein himself \cite{Bek3}
  gave a prescription for a cosmological extension
by choosing $R$ in Eq. (\ref{BEB}) to be the particle horizon.
Amusingly, he arrived at the conclusion that
the  bound is violated sufficiently near the big-bang singularity, implying
that the latter is fake (if the bound is always valid).
About a year ago,
 Fischler and Susskind
(FS) \cite{FS} proposed a similar extension of the
holographic bound to cosmology, arguing that the area of the particle
horizon should bound  entropy on the backward-looking
light cone according to (\ref{HOL}). It was soon realized, however, that
the FS proposal requires modifications, since violations of it were
found to occur in physically reasonable situations.
An improvement of the FS bound applicable to light-like hypersurfaces
was later made by Bousso \cite{Bousso}.

  Of more interest here are the attempts  made at
 deriving cosmological entropy bounds on space-like hypersurfaces
 \cite{EL,GV2,BR,KL,BV3}.
These identify the maximal size of a
spatial region for which holography works:  the Hubble radius  
\cite{EL, GV2, KL},
the apparent horizon \cite{BR}, or, finally, a so-called causal
connection (Jeans) scale \cite{BV3}.

For our purposes here, we do not need to enter into the relative merits
of these various proposals. Rather,
we will only outline the physical idea behind them.
Consider, inside a quasi-homogeneous Universe, a sphere of radius $ H^{-1}$.
We may consider ``isolated" bodies, in the sense of ref. \cite{Bek1},  
fully contained in
the sphere, i.e. with radius $R < H^{-1}$. For such systems,
 the usual BB holds and is saturated by a black hole of  size $R$.
We may consider next several black holes inside our Hubble volume,  
each carrying an entropy
proportional to the square of its mass. If two, or more, of these  
black holes merge,
their masses will add up, while the total entropy after the
merging,  being quadratic in the total mass, will exceed the sum of  
the initial entropies.
In other words, in order to maximize entropy, it pays to form  black holes 
as large as possible.

Is there a limit to this process of entropy increase?
The suggestion made in \cite{EL,GV2,BR,KL,BV3}, which finds support in
old results by several groups \cite{Carr}, is that a critical length  
of order $H^{-1}$
is the upper limit on how large a classically stable black hole can be.
If we accept this hypothesis, the upper bound on the entropy contained  
in a given region ${\cal R}$ of space
will be given by the number of Hubble volumes in ${\cal R}$, $n_H = V H^3$
times the Bekenstein--Hawking entropy of a BH or radius $H^{-1}$,
$H^{-2} l_P^{-2}$. The two factors can be combined in the suggestive formula:
\begin{equation}
S({\cal R}) < l_P^{-2} \int_{{\cal R}}  d^3 x ~ \sqrt{h} \tilde{H}  
\equiv S_{HB} \;,
\label{HEB}
\end{equation}
where $\int_{{\cal R}}  d^3 x ~ \sqrt{h}$ is the volume of the  
space-like hypersurface whose entropy we wish to bound, and $\tilde{H}$  
differs from one proposal to another \cite{EL,GV2,BR,KL,BV3}, but is,  
roughly, of the order of
 the Hubble parameter. Actually, since $H$ is proportional
to the trace of the second fundamental form on the hypersurface, Eq.  
(\ref{HEB}) reminds us of the
boundary term that has to be added to the gravitational action in order to
correctly derive Einstein's equations from the usual variational principle.
This shows that the bound (\ref{HEB}) is generally covariant for  
$\tilde{H}=H$. It can also be written covariantly for the  
identification of $\tilde{H}$ made in \cite{BV3}.

For the qualitative discussion that follows, let us simply take  
$\tilde{H} = H$ and let
 us convert the bound to string-frame quantities, taking into account  
the relation between
$l_P$ and $\lambda_s$, given in Eq. (\ref{VEV}). We obtain \cite{GV2}:
\begin{equation}
S({\cal R}) <  (VH^3) (H^{-2} \lambda_s^{-2} e^{-\phi}) =  
e^{-\bar{\phi}} H \lambda_s^{-2}  \;,
\label{HEBSF}
\end{equation}
where we have fixed an  arbitrary additive constant in
 the definition  (\ref{sdildef}) of $\bar{\phi}$. Equation  
(\ref{HEBSF}) thus connects very simply
 the entropy bound of a region of fixed comoving volume to the most important
variables occurring in string cosmology (see, e.g., the phase diagram  
of Fig. 3).

An immediate application of the bound (\ref{HEBSF}) was pointed out in  
\cite{GV2}.
Noting that the bound is initially saturated in the BDV picture \cite{BDV}
 of collapse/inflation, the bound itself cannot decrease without a  
violation of the second law.
This gives immediately:
\begin{equation}
\dot{\bar{\phi}} \le \frac {\dot{H}}{H}  \;.
\label{HEBcond}
\end{equation}
It is easy to check that this inequality is  obeyed, but just so,
 during DDI, in the sense that
it holds with the equality sign. In other words, the HEB is saturated  
initially {\it and
throughout} DDI in the BDV picture.
The bound also turns out to give a physically
acceptable value for the entropy of the Universe just after the big  
bang: a large entropy
on the one hand (about $10^{90}$); a small entropy for the total mass  
and size of the observable
Universe on the other, as often pointed
out by Penrose \cite{PenEntr}. Thus, PBB cosmology neatly explains why the
 Universe, at the big bang,
{\it looks} so fine-tuned (without being so) and provides a natural  
arrow of time in the direction
of higher entropy \cite{GV2}.

What happens in the mysterious string phase,
where we are desperately short of reliable techniques?
It is quite clear that Eq. (\ref{HEBcond}) does not allow $H$ to reach
saturation ($\dot{H} =0$)  in the first quadrant
of Fig. 3 since  $\dot{\bar{\phi}}>0$ there. Instead, saturation of  
$H$ in the second quadrant
(where $\dot{\bar{\phi}} \le 0$) is perfectly all right. But this implies
having attained the sought for branch change!

Let us finally look at the loop corrections. Since, physically, these  
correspond to
taking into account the back-reaction from particle production, we may  
check when
the entropy in the cosmologically produced particles
starts to threaten the bound. As discussed in Subsection 4.6, the  
entropy density
in quantum fluctuations is given by $ \sigma \sim N_{eff} H^3$, which
equals the bound $\sigma_{HEB} \sim H l_P^{-2}$ precisely when $l_P^2  
H^2 N_{eff} = O(1)$.
But, as already pointed out, this is just the  line on which
the energy density in quantum fluctuations becomes critical (see Eq.  
(\ref{PBBB})) and where,
according to \cite{GMV}, the back-reaction becomes $O(1)$.
Similar conclusions are reached by applying generalized second law  
arguments \cite{RB}.

The picture that finally emerges from all these considerations is best  
illustrated
with reference to the diagram of  Fig. 4.
 Two  lines are shown,  representing
boundaries for the possible evolution. The horizontal boundary is  
forced upon by
the large-curvature corrections, while the tilted line in the first  
quadrant corresponds to
the equation $l_P^2 H^2 N_{eff} = O(1)$ that
we have just discussed. Amusingly, this line
was also suggested by Maggiore and Riotto \cite{MR} as a boundary  
beyond which
copious production of $0$-branes would set in.
Thus, depending on initial conditions, the PBB bubble corresponding to  
our Universe
would hit first either the high-curvature or the large-entropy boundary and
initiate an exit phase. Hopefully, a universal late-time attractor  
will emerge
guiding the evolution into the FRW phase of standard cosmology
(shown as a vertical line in Fig. 10).

Needless to say, all this has to be considered, at best, as having  
heuristic value.
If taken seriously, it would suggest that the Universe will never enter
the strong-coupling, strong-curvature regime, where the largely
unknown M-theory should be used.
The low-energy limit of the latter (the much better understood 11-D  
supergravity)
could suffice to deal with the fundamental exit problem of string cosmology.
We refer to the literature for several other attempts at M-cosmology  
\cite{branes}.

\vskip 5mm

\section{Outlook}
The outlook for the pre-big bang scenario,  as  formulated at present,  
is not necessarily an optimistic one.
I am not sure I would bet a lot of money on it being right!
But this is not really the issue. We have to remember that the PBB  
scenario is
 a top--down approach to cosmology. As stressed in the introduction,
it would be quite a miracle if the correct model could be guessed without
extensive feed-back from the data. The good news here is that new data  
are coming in all the time, and will continue
 to do so with more and more precision in the coming years!

Rather, we should draw some lessons from this new attempt at very  
early cosmology, whether
 it succeeds or it fails. As I can see, the main lessons to be drawn  
are the following:
\begin{itemize}
\item  Our Universe did not have to emerge, together with space and  
time, from a singularity;
in string theory, the singularity should be fake, either because
 it is tamed by finite-string-size effects, or
because it simply signals the need for new degrees of freedom in the  
description of physics
at very short distances;
\item Because string theory is an extension of GR, inflation is possible 
in that context even in the absence of potential
energy (i.e. of an effective cosmological constant);
actually, inflation is very natural and easy to achieve,
being a consequence of the
duality symmetries of the string-cosmology equations;
\item Inflation in string cosmology can be related, mathematically,
to the problem of gravitational collapse in GR; as such, it is a  
generic phenomenon, once
the assumption of asymptotic past triviality is made;
furthermore, the curvature scale and the coupling
at the onset of PBB inflation are arbitrary classical moduli;
\item The Universe did not have to start hot!
A hot Universe can emerge from a cold one thanks to quantum particle  
production
in inflationary backgrounds;
\item PBB cosmology predicts a rich spectrum of perturbations with  
different spectra
depending on each perturbation's ``pump" field and on its evolution in the
PBB era; observable relics of these perturbations may serve as a window
on physics in the pre-bangian Universe all the way down to
the string/Planck scale;
\item The simplest PBB models either predict too small perturbations  
at large scales,
or a spectrum of isocurvature perturbations which may be already  
``experimentally
challenged" (as Rocky Kolb would kindly say);
\item The exit problem still remains the hardest theoretical challenge
to the whole idea of  PBB cosmology;
\item Hopefully,  the combination of the above-mentioned experimental  
and theoretical challenges will be able to
tell us whether the PBB idea is just doomed, or whether parts
 of it should be kept while searching
for a better scenario; it should also suggest new avenues for physics-driven
research in string/M-theory;
\item Last but not least, the PBB idea has taught us that we need not  
lock ourselves
into preconceived ideas in cosmology (cf. ``the big bang is the  
beginning of time", ``inflation needs a scalar potential"); rather, we  
should contemplate as wide a range of theoretically sound
possibilities as we can in order for Nature to choose, at best, one of them.

\end{itemize}

\vskip 5mm

\centerline {ACKNOWLEDGEMENTS}

I am very grateful to Pierre Bin\'etruy and Richard Schaeffer for having  
invited me to such
a pleasant and stimulating school, to Fran\c cois David for the perfect
organization, and to all the students for their patience in listening
(after 5 weeks of courses!) and for their interesting questions.

\end{document}